\documentclass[10pt]{article}
\usepackage{amsmath}
\usepackage{amssymb}
\usepackage{amsthm}
\usepackage{color}
\usepackage[margin=0.75in]{geometry}
\usepackage{graphicx}
\usepackage[citestyle=numeric, bibstyle=numeric, natbib=true, backend=bibtex]{biblatex}
\usepackage{hyperref}
\usepackage{subcaption}
\usepackage{authblk}
\usepackage{booktabs}

\def\E{\mathbb{E}}
\def\P{\mathbb{P}}
\def\R{\mathbb{R}}

\newtheorem{theorem}{Theorem}[section]

\newtheorem{lemma}[theorem]{Lemma}

\newtheorem{definition}[theorem]{Definition}

\title{Cluster Randomized Designs for One-Sided Bipartite Experiments}
\author[1]{Jennifer Brennan\thanks{Work done while the author was an intern at Google Research. Correspondence to \texttt{jrbrennan@google.com}.}}
\author[2]{Vahab Mirrokni}
\author[2]{Jean Pouget-Abadie}

\affil[1]{
  Paul G. Allen School of Computer Science \& Engineering,
  University of Washington,
  Seattle, WA}
\affil[2]{
   Google Research,
   New York, NY}

\begin{document}

\maketitle

\begin{abstract}
The conclusions of randomized controlled trials may be biased when the outcome of one unit depends on the treatment status of other units, a problem known as \textit{interference}.
In this work, we study interference in the setting of one-sided bipartite experiments in which the experimental units---where treatments are randomized and outcomes are measured---do not interact directly.
Instead, their interactions are mediated through their connections to \textit{interference units} on the other side of the graph.
Examples of this type of interference are common in marketplaces and two-sided platforms.
The \textit{cluster-randomized design} is a popular method to mitigate interference when the graph is known, but it has not been well-studied in the one-sided bipartite experiment setting.
In this work, we formalize a natural model for interference in one-sided bipartite experiments using the exposure mapping framework.
We first exhibit settings under which existing cluster-randomized designs fail to properly mitigate interference under this model. We then show that minimizing the bias of the difference-in-means estimator under our model results in a balanced partitioning clustering objective with a natural interpretation.
We further prove that our design is minimax optimal over the class of linear potential outcomes models with bounded interference.
We conclude by providing theoretical and experimental evidence of the robustness of our design to a variety of interference graphs and potential outcomes models.
\end{abstract}

\section{Introduction}\label{sec:intro}

Interference is a well-studied phenomenon in causal inference, whereby the treatment status of one unit can affect the outcome of another. Formally a violation of the Stable Unit Treatment Value Assumption~\cite{rubin1980discussion}, interference has been studied in many settings, including agricultural studies~\cite{neyman1923application}, clinical trials~\cite{hudgens2008toward}, social networks~\cite[chap.~16 of][]{aronow2021spillover, aral2011creating, gui2015network, eckles2016estimating}, and marketplaces~\cite{blake2014marketplace, rolnick2019randomized, li2022interference, wager2021experimenting, munro2021treatment}. 
Marketplaces exhibit unique forms of interference because they involve two types of units: \textit{buyers} and \textit{sellers}. Units of the same type do not interact directly, but rather their interactions are mediated through their interactions with units of the opposite type. For example, when buyers compete to buy limited goods, an increase in the price one buyer is willing to pay for a good will affect the market-clearing rate for that good, thus increasing the price for all other buyers.

Marketplace experiments can be conceptualized as running an experiment on a bipartite graph
between buyers and sellers, with edges representing buyer-seller interactions.
Other settings beyond marketplaces can also be formalized as bipartite graphs, such as  content platforms (matching viewers to creators) or ride-sharing apps (matching riders to drivers) \cite{chamandy2016experimentation}.
We consider perhaps the most straightforward way of running an experiment on a bipartite graph: by treating and measuring a single side of the graph, a setting we call the \emph{one-sided bipartite experiment framework}. We refer to the units on this side of the graph as \emph{experimental units} since they receive treatment and their outcomes appear in our estimators. We refer to the units on the other side of the graph as \emph{interference units}: while they do not explicitly receive treatment, they mediate the interactions between experimental units.

The presence of interference units distinguishes one-sided bipartite experiments from other settings of network interference because the interactions between experimental units are known in the latter case, but must be inferred from the interactions with interference units in the former. In this work, we show that a popular class of experimental designs for network interference, \textit{cluster-randomized designs}, does not extend immediately to this setting.
Instead, we propose a variant of the cluster-randomized design, with a clustering objective that accounts for the way in which interference units mediate the interference between experimental units. We motivate the use of the difference-in-means estimator in this setting and show that clustering according to our objective minimizes the bias of that estimator in a minimax sense over the class of linear potential outcomes models with bounded interference. 
We further illustrate the failure of existing clustering designs when applied to one-sided bipartite experiments, and conclude by illustrating the robustness of our design to a variety of potential outcomes models and bipartite graphs both theoretically and empirically.

\subsection{Related work}\label{sec:relatedWork}

Accurate estimation of the treatment effect requires some knowledge of the mechanism of interference \cite{basse2018limitations}; otherwise, having even a single treated (resp. control) unit in a graph of control (resp. treated) units could change the outcome of every unit arbitrarily. Existing work varies in the strength and nature of the assumptions on the interference model, but in general these assumptions take one of two types. The \textit{exposure mapping} approach \cite{hudgens2008toward, manski2013identification}, formalized by \citet{aronow2017estimating}, defines a notion of when a unit is ``completely treated'' or ``completely controlled,'' then uses the inverse propensity score (IPS, also known as Horvitz-Thompson) estimator to construct an unbiased estimate of the average treatment effect. By contrast, an alternate approach is to propose a model for the effect of interference on the potential outcomes, and then rely on this model to estimate the average treatment effect using data from all units (even the ones experiencing a great deal of interference) \cite{gui2015network, eckles2017design, pouget2019variance, harshaw2022design}.
Since the quality of the estimate depends on the accuracy of the modeling assumptions, several methods have been developed to estimate the magnitude and form of interference \cite{aral2011creating, saveski2017detecting, aronow2012general, athey2018exact, toulis2013estimation, savje2021average}.
\citet{chin2019regression} lessens the reliance on the form of the interference model by developing a model-agnostic regression estimator that reduces average treatment effect estimation to a problem of feature engineering; however, it assumes that each unit's response follows a shared (but unknown) model, which is more restrictive than the potential outcomes model we use in this work.

Even in cases when the model is only approximate, so that including data from units experiencing interference may bias the estimator, \citet{eckles2017design} use realistic graph models to argue that the bias incurred is more than offset by the reduction in variance achieved by avoiding an IPS estimator.
Our work assumes this same regime, in which the graphs are sufficiently dense that an IPS estimator will be too high-variance to be practical, so we accept an estimator with some bias. We assume a linear model of the potential outcome on the measured exposure, as done in \cite{gui2015network, harshaw2022design}. To avoid strong dependence on the linear assumption, we use the difference-in-means estimator which does not rely on the model of interference, and we choose our experimental design to be minimax optimal over the class of linear potential outcomes models.

Given some model of interference and a choice of estimator, the next question is how to design an experiment (an assignment of units to treatment or control).
The most popular experimental design in the case of network interference is the cluster-randomized design, studied by \cite{ugander2013graph, eckles2017design, candogan2021near} in the case of non-bipartite graphs. This design first clusters units according to the provided graph, and then assigns each cluster to either treatment or control.
In the specific case of bipartite graphs, other works propose modifications to unit-level randomization by choosing which side of the graph to randomize \cite{johari2022experimental, li2022interference, bajari2021multiple}. The authors of \cite{pouget2019variance, harshaw2022design} look directly at clustered designs on bipartite graphs, but they study a two-sided experimental framework in which one side of the graph is randomized while the other is measured. 
Perhaps closest to our work is that of \citet{rolnick2019randomized}, which suggests a balanced partitioning of geographical regions using a clustering objective that is similar to ours. Their work considers a more restrictive form of the potential outcomes model, and uses clustering heuristics that are specific to the geographical setting. By contrast, our work considers the robustness of the design to a broader class of potential outcomes models and extends beyond geographical regions to general interference graphs.

\section{Models and estimators}

We now formalize the interference model using the potential outcomes framework~\cite{imbens2015causal}. Let $\mathbf{Z}\in\{-1,1\}^N$ be an assignment of each of the $N$ experimental units to treatment ($Z=1$) or control ($Z=-1$). The \textit{potential outcome} of the $i^{th}$ experimental unit is denoted $Y_i$, and in the most general setting could be a function $Y_i(\mathbf{Z})$ of the entire treatment assignment vector. We further assume the existence of a (known) bipartite graph between experimental units and interference units, with nonnegative weights $w_{is}\geq 0$ encoding the relationship between experimental unit $i$ and interference unit $s$. Such a graph may be obtained from historical data on interactions between units $i$ and $s$, or on similarity between $i$ and $s$ measured by geography or other features \cite{rolnick2019randomized, zigler2021bipartite}.

While there are many possible estimands of interest, our primary goal is to estimate the \textit{average total treatment effect} $\tau = \frac{1}{N} \sum_{i\in [N]} Y_i(\mathbf{Z}=\mathbf{1}) - Y_i(\mathbf{Z}=-\mathbf{1})$, sometimes referred to as the average treatment effect or total treatment effect. Since we assign some units to treatment and others to control, it is impossible to observe any potential outcome under a fully treated $(\mathbf{Z}=\mathbf{1})$ or fully controlled ($\mathbf{Z}=-\mathbf{1}$) condition. 
If the underlying interference graph among units were composed of multiple connected components, so that only the assignments $Z$ in a unit's connected component affected its potential outcome, then it would be possible to assign treatment at the level of the connected component and observe fully treated or control outcomes. However, in realistic marketplaces, such a perfect separation almost never occurs.
As a result, we require some modeling assumptions on the potential outcomes 
to infer the behavior of unit $i$ under the fully treated or controlled condition.

\subsection{The potential outcome model}

A popular approach to modeling potential outcomes with network interference is the \textit{exposure mapping} paradigm \cite{hudgens2008toward, manski2013identification, aronow2017estimating}. An exposure mapping is a function $e_i: \{-1,1\}^N \to \R$ such that $Y_i(Z_i, e_i(\mathbf{Z})) = Y_i(\mathbf{Z})$. In other words, the indirect effect of $Z_{j\neq i}$ on unit $i$ is captured completely by the exposure $e_i(\mathbf{Z})$. When it is clear from context, we will write $e_i$ instead of $e_i(\mathbf{Z})$ to denote the exposure of unit $i$. Given an exposure mapping, we can further posit a model for the effect of assignment $Z_i$ and exposure $e_i$ on the outcome $Y_i$. 
We principally consider the linear model
\begin{align}
    Y_i(\mathbf{Z}) &= \alpha_i + \beta_i Z_i + \gamma_i e_i,\label{eqn:linearPotentialOutcomes}
\end{align}
which is commonly used in the interference literature \cite{gui2015network, harshaw2022design}, 
although other models are discussed in Section \ref{sec:robust}.
Following the tradition of the finite population model, first defined by \citet{neyman1923application}, we treat the coefficients $\alpha_i$, $\beta_i$ and $\gamma_i$ as fixed but unknown so that the only randomness in the observation model is due to the choice of treatment assignment $\mathbf{Z}$. This contrasts with alternatives such as a model in which coefficients are drawn from some super-population or a model in which the coefficients are common across $i$ with some additive error term $\varepsilon$ in the linear model \cite{chin2019regression}. The finite population model avoids making assumptions about a population from which experimental units are drawn, and for this reason is often preferred in the causal inference literature \cite{rubin1990comment, ding2017bridging}.

\begin{figure}
    \centering
    \includegraphics[clip, trim=0cm 6cm 0cm 0cm, width=\textwidth]{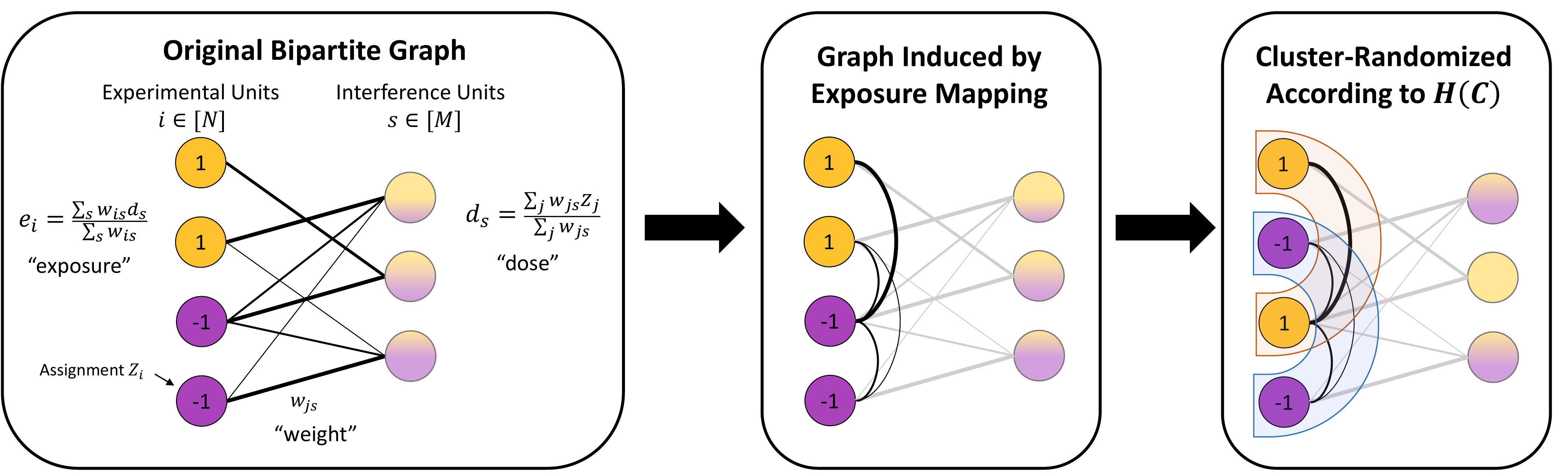}
    \caption[Overview of our cluster-randomized experimental design]{Overview of our cluster-randomized experimental design. Left Panel: We are given a bipartite graph connecting experimental units (left) with interference units (right) with edges of known weight $w$. Assigning treatments $Z_i\in\{-1,1\}$ to each experimental unit induces a dose $d_s$ on each interference unit, and an exposure $e_i$ on each experimental unit. Middle Panel: The exposure of experimental unit $i$ depends on the assignment of other experimental units $j$, which induces a graph on the experimental units. Right Panel: Clustering the experimental units based on the induced graph creates clusters of units which heavily influence each others' outcomes. Randomizing the treatment assignment at the level of the cluster provides exposures $e_i$ that are much closer to the treatment $Z_i$ than would be achieved with a unit-randomized design.}
    \label{fig:bipartiteFig}
\end{figure}

\subsection{The exposure model}
The literature on network interference typically defines the exposure $e_i$ as a function of the treatments of the neighbors of $i$ \cite{hudgens2008toward, manski2013identification, aronow2017estimating}. For example, the exposure mapping might be the (weighted) fraction of neighbors that are treated \cite{eckles2017design}, or the count of neighbors that are treated \cite{ugander2013graph}. In the bipartite setting, experimental units are never immediate neighbors. Instead, experimental units interact through their relationships with interference units. Defining an exposure mapping that is analogous to those used in the network setting requires an analogous definition of the ``neighborhood'' of an experimental unit, as well as the weight of the connections from a unit to each of its neighbors.

We propose a bipartite analogue to the neighborhood-based exposure mapping, composed of two parts: the \textit{dose} $d_s\in[-1,1]$ of each interference unit $s\in[M]$ represents the weighted average of the treatment assignment $Z_i$ of each experimental unit in the neighborhood of $s$, while the \textit{exposure} $e_i\in[-1,1]$ of each experimental unit $i\in[N]$ represents the weighted average of doses among the interference units in the neighborhood of $i$:
\begin{align}
    d_s = \frac{1}{\sum_{i\in[N]} w_{is}} \sum_{i\in[N]}w_{is} Z_i,
    &\qquad \qquad
    e_i = \frac{1}{\sum_{s\in[M]} w_{is}} \sum_{s\in[M]}w_{is} d_s.\label{eqn:exposureMappingDfn}
\end{align}
See Figure \ref{fig:bipartiteFig} for an illustration of the exposure mapping.
Because the exposure thus defined can be written as a linear combination of the treatment assignments of the two-hop neighbors of unit $i$, our definition of exposure can be viewed as an exposure mapping where $e_i$ depends on two-hop neighbors, in which we must impute the effective ``assignment'' of interference units $s$ from the assignments of their neighbors.
Depending on the problem instance,
it may be more appropriate to write one or both of these terms as an unnormalized linear combination instead of a convex combination; we show in Appendix \ref{appendix:normalization} that all of our results apply to the unnormalized setting as well.

\subsection{Estimators}\label{sec:estimators}

Estimators for the average total treatment effect $\tau$ under network interference typically fall into one of two categories: difference-in-means (DIM) estimators and inverse propensity score (IPS) estimators. The former reports the difference between the mean of $N_T$ treated units and the mean of $N_C$ control units, where we treat $N_T$ and $N_C$ as fixed quantities chosen before treatment randomization occurs. The latter requires a notion of which units are ``fully exposed'' to treatment or to control, perhaps defined by $e_i$, and reweighs fully exposed observations by the inverse probability of achieving that state.
\begin{align*}
    \widehat\tau_{DIM} = \sum_{Z_i=1} \frac{Y_i}{N_T} -\sum_{Z_i=-1} \frac{Y_i}{N_C},
    &\quad
    \widehat\tau_{IPS} = \frac{1}{N}\sum_{i\in[N]} \frac{Y_i\mathbf{1}\{i~\text{fully treated}\}}{\P(i~\text{fully treated})} - \frac{Y_i\mathbf{1}\{i~\text{fully controlled}\}}{\P(i~\text{fully controlled})}
\end{align*}
The difference in means estimator can suffer from bias in the presence of interference, because both the treated and control means may be biased estimates for their population quantities $n^{-1}\sum_i Y_i(\mathbf{1})$ and $n^{-1}\sum_i Y_i(-\mathbf{1})$, respectively. The IPS estimator is unbiased as long as a ``fully treated'' unit does in fact behave like a unit in which $\mathbf{Z}=\mathbf{1}$ (and likewise for control), however, this comes at a cost of high variance if the propensity scores are small. In our linear model, full exposure only occurs when $e_i=Z_i$, which requires all experimental units in the two-hop neighborhood of $i$ to have the same assignment as $i$. In many realistic bipartite graphs the chance of full exposure for a given unit is very low, increasing the bias in the IPS estimator to unacceptable levels. Appendix \ref{appendix:ips} illustrates the high variance of the IPS estimator for a selection of simulated bipartite graphs and a variety of definitions of ``full treatment.'' The difference-in-means estimator also has the advantage that it outperforms the IPS estimator in the setting of no interference, $Y_i(\mathbf{Z}) = Y_i(Z_i)$, since in this setting $\widehat\tau_{DIM}$ is unbiased and has the same or lower variance than $\widehat\tau_{IPS}$ for any definition of full exposure. As a result, the difference in means estimator can be seen as an ``optimistic'' choice of estimator, which makes better use of the data in the case of no interference while incurring bias in the presence of interference.
For the reasons described above, we focus on the difference in means estimator in this work.
\section{Experimental design}\label{sec:experimentalDesign}

Having chosen the linear potential outcomes model of Equation \eqref{eqn:linearPotentialOutcomes} with the exposure mapping of Equation \eqref{eqn:exposureMappingDfn} and the difference-in-means estimator $\widehat\tau_{DIM}$, we seek a mechanism for randomly assigning experimental units to treatment or control 
that achieves low mean squared error (MSE) in recovering the average total treatment effect $\tau$. The MSE can be decomposed into the bias and the variance, where the former is caused by interference and the latter is primarily a function of the number of units randomized to treatment and control. 
Following previous works \cite{ugander2013graph, eckles2017design, candogan2021near}, we consider the class of \textit{balanced cluster-randomized designs},
formalized in Definition \ref{def:balancedDesign}.
\begin{definition}[Balanced $K$-cluster randomized design]\label{def:balancedDesign}
Let $\mathcal{C}=\{C_\ell\}_{\ell=1}^K$ be a partition of the $N$ experimental units into $K$ equally sized clusters.
A balanced $K$-cluster randomized design $\mathcal{D}(\mathcal{C})$ is a distribution over vectors $\mathbf{Z}\in\{-1,1\}^N$ generated by assigning $Z_i = 1$ for all experimental units $i$ belonging to a set of $K_T\in(0,K)$ clusters chosen uniformly at random among all $K$ clusters.
\end{definition}
If the clustering faithfully captures patterns of interference, cluster randomization can increase the chance that a treated (resp. control) unit is highly exposed to treatment (resp. control), thereby reducing the bias in $\widehat\tau_{DIM}$. 

We control the variance of the estimator by enforcing a balanced clustering, so that assigning a fixed fraction of clusters to treatment results in a consistent fraction of units assigned to treatment. 
Since cluster randomization can change the effective number of experimental units \cite[Section 4.2]{rolnick2019randomized}, the variance of $\widehat{\tau}_{DIM}$ will depend on the number of clusters chosen. In practice, the variance under a given clustering can typically be estimated using historical data, often called an A/A test. Practitioners can use these estimates to control the variance in light of the anticipated effect sizes. The bias due to interference, however, is not estimable with historical data when all such data is observed under the control condition. 
Balanced clustering also has practical implementation benefits, which we discuss further in Appendix \ref{sec:future}.

Since the variance can be estimated from historical data and controlled by the number of balanced clusters, we choose to focus on identifying a clustering that minimizes the bias of $\widehat\tau_{DIM}$ given a fixed number of balanced clusters. Choosing the correct number of clusters to trade off bias and variance remains an important direction for future work.

\subsection{Existing designs do not work in the one-sided bipartite setting}\label{sec:design:existing}

A natural question is whether existing approaches to cluster-randomized designs are adequate for our setting of one-sided experiments on bipartite graphs.
In this section, we consider the natural extensions of two clustering algorithms from the literature to our setting and describe failure modes for each of them.

\textbf{Direct clustering of the bipartite graph.} 
Cluster-randomized designs have primarily been studied in the context of graphs in which all units are experimental units, such as social networks \cite{ugander2013graph, eckles2017design}. 
Within the class of cluster-randomized designs, balanced partitioning has been suggested as a means to minimize bias \cite{eckles2017design}.
The natural analog in the bipartite setting is to create a balanced partitioning of the bipartite graph according to the provided edge weights $w_{is}$; this extension was considered as a baseline in a different bipartite setting by \citet{pouget2019variance}. Clusters may contain both experimental units and interference units, but only experimental units count toward the balancedness constraint.

Unfortunately, this naive approach fails to take into account the two-hop structure of interference in a bipartite graph. Consider for example two experimental units $i$ and $j$ that must be assigned to a different cluster than that of their common neighbor, interference unit $s$, perhaps to satisfy a balancedness or cluster cardinality constraint. There is no benefit for the balanced partitioning algorithm on the bipartite graph to assign these two experimental units to the same cluster as each other, despite them sharing a common neighbor. This mode of failure is illustrated with an example in Appendix \ref{appendix:design:direct}.

\textbf{Maximizing the variance of the doses.} 
Cluster-randomized designs have also been considered in the setting of two-sided bipartite experiments, in which treatment is assigned to the experimental units but outcomes are measured on the interference units. 
When the potential outcome of an interference unit is a linear function of the dose $d_s$, \citet{pouget2019variance} and \citet{harshaw2022design} 
recommend assigning treatments $\mathbf{Z}$ in a way that maximizes the empirical variance of the realized doses. In the one-sided bipartite setting it is important to enforce balancedness of the clusters to control the variance of $\widehat\tau_{DIM}$, so an extension of their objective to this setting would be to maximize $\text{Tr}(\text{Var}(\mathbf{d}))$ over a balanced cluster-randomized assignment. 

Unfortunately, this clustering objective fails in the one-sided bipartite setting. It is possible that the doses $d_s$ of the interference units are primarily controlled by a small number of experimental units with especially large weights. In this case, ensuring that the doses are close to $-1$ or $1$ only translates into guarantees about those highly-weighted experimental units, and not necessarily about the typical experimental unit. This idea is illustrated with an example in Appendix \ref{appendix:design:varmax}.

\subsection{The bias-minimizing clustering objective $\mathcal{H}(\mathcal{C})$}
Having shown that these two natural extensions fail to identify the bias-minimizing cluster-randomized designs, we turn our attention to a clustering objective $\mathcal{H}(\mathcal{C})$ that directly minimizes the bias in $\widehat\tau_{DIM}$. 

\begin{lemma}\label{lem:dim-bias}
Suppose the difference in means estimator $\widehat\tau_{DIM}$ is computed on a balanced $K$-cluster design $\mathcal{D}(\mathcal{C})$.
Let the potential outcomes follow the linear model \eqref{eqn:linearPotentialOutcomes}. Then the bias is given by:
\begin{align*}
    \tau - \E_{\mathbf{Z}\sim \mathcal{D}(\mathcal{C})}[\widehat\tau_{DIM}] &=
    \frac{2}{N} \cdot  \frac{K}{K-1} \sum_{i\in[N]} \sum_{j\not\in\mathcal{C}(i)} \gamma_i \sum_{s\in[M]} \frac{w_{is}}{\sum_s w_{is}} \frac{w_{js}}{\sum_k w_{ks}}.
\end{align*}
Suppose further that all interference terms are bounded in magnitude by a constant, so that $\gamma_i=O(1)$. Then the minimax bias is bounded:
\begin{align*}
    \arg\min_\mathcal{C} \max_{\mathbf{\gamma}:~\gamma_i=O(1)} \left|\tau - \E_{\mathbf{Z}\sim \mathcal{D}(\mathcal{C})}[\widehat\tau_{DIM}] \right|
    &= \arg\min_{\mathcal{C}} \sum_{i\in[N]} \sum_{j\not\in\mathcal{C}(i)} \sum_{s\in[M]} \frac{w_{is}}{\sum_s w_{is}} \frac{w_{js}}{\sum_k w_{ks}} \\
    &=: \arg\min_{\mathcal{C}}\mathcal{H}(\mathcal{C}).
\end{align*}
\end{lemma}
To interpret Lemma \ref{lem:dim-bias}, we can consider the cases in which the clustering $\mathcal{C}$ can achieve zero bias for $\widehat{\tau}_{DIM}$. One way this can occur is when there is no interference, so that $\gamma_i = 0$ for all $i$. Even in the presence of interference, the bias of $\widehat{\tau}_{DIM}$ can be zero if the cut edges $w_{is}~:~j\not\in\mathcal{C}(i)$ are all of zero weight. In such a well-clustered graph, the outcome of unit $i$ depends only on the treatment assignments of other units within its own cluster. Since the same treatment is applied to each element of a given cluster, a perfectly clustered graph has $e_i=Z_i$ and experiences no bias due to interference.

Our objective $\mathcal{H}(\mathcal{C})$ has two interpretations: as a clustering objective on an induced graph on experimental units, and as a statistical objective on the covariance between assignments and exposures.

\textbf{A graphical interpretation of $\mathcal{H}(\mathcal{C})$.}
Interference occurs when the treatment assignment of unit $j$ affects the exposure of unit $i$. The influence of $Z_j$ on $e_i$ under the linear interference model \eqref{eqn:linearPotentialOutcomes} is
\begin{align}
    e_i(\mathbf{Z}_{j^+}) - e_i(\mathbf{Z}_{j^-})
    &= 2 \sum_{s\in[M]} \frac{w_{is}}{\sum_s w_{is}} \frac{w_{js}}{\sum_k w_{ks}}.\label{eq:graphicalInterpretation}
\end{align}
where $\mathbf{Z}_{j^+}$ (resp. $\mathbf{Z}_{j^-}$) is the vector $\mathbf{Z}$ with entry $Z_j$ set to 1 (resp. $-1$).
Now consider a directed graph on experimental units where the edge from $i$ to $j$ is weighted according to $e_i(\mathbf{Z}_{j^+}) - e_i(\mathbf{Z}_{j^-})$; clustering according to this graph minimizes the objective $\mathcal{H}(\mathcal{C})$. This is in fact a similar approach that was adopted in \cite{rolnick2019randomized}, which also looked at a similar folding of the graph and showed it to minimize the bias under a linear potential outcome model motivated by geographical migrations.

\textbf{A statistical interpretation of $\mathcal{H}(\mathcal{C})$.}
A natural goal of a cluster-randomized design is to ensure that the exposure $e_i$ is close to the treatment $Z_i$. Lemma \ref{lem:covarianceObjective} shows that our clustering objective $\mathcal{H}(\mathcal{C})$ maximizes the covariance between the exposures and assignments.
\begin{lemma}\label{lem:covarianceObjective}
Let $\mathcal{D}(\mathcal{C})$ be a balanced $K$-cluster randomized design. Then we have: 
$\arg\max_{\mathcal{C}} \text{Tr}\left( \text{Cov}_{\mathbf{Z} \sim \mathcal{D}(\mathcal{C})}(\mathbf{Z}, \mathbf{e}) \right)
    = \arg\min_{\mathcal{C}} \mathcal{H}(\mathcal{C}).$
\end{lemma}
Finally, we observe that our cluster-randomized design is an instance of the well-known \textit{balanced partitioning} problem~\cite{andreev2006balanced} on the graph over experimental units, with edges given by the explicit formula \eqref{eq:graphicalInterpretation}. Balanced partitioning is NP-hard, even in its relaxed form which enforces only partial balancedness, but many tools exist to compute it approximately \cite{andreev2006balanced, aydin2019distributed, nishimura2013restreaming}.

\section{Robustness}\label{sec:robust}
So far we have shown that a cluster-randomized design with objective $\mathcal{H}(\mathcal{C})$ minimizes the bias of $\widehat\tau_{DIM}$ under the normalized exposure mapping in by Equation \eqref{eqn:exposureMappingDfn} and the linear potential outcomes model in Equation \eqref{eqn:linearPotentialOutcomes}. A natural question is: how robust is this design to deviations from this model? 
In this section we analyze the robustness of the design under alternative potential outcomes models, showing that the design remains minimax optimal for Lipschitz potential outcomes and minimizes an upper bound on the bias for a class of potential outcomes models motivated by the exposure mapping literature. Additionally, there may exist experimental settings where the exposure $e_i$ is better modeled without the normalization constant $\sum_s w_{is}$, such as when the outcome $Y_i$ is proportional to the edge weight incident to unit $i$. In this case, or in the case that the dose $d_s$ is unnormalized, all of our results remain valid for a generalization of the clustering objective which removes the corresponding normalization constant(s) from $\mathcal{H}(\mathcal{C})$. This idea is formalized in Appendix \ref{appendix:normalization}.

\subsection{Lipschitz potential outcomes}
Let $Y_i(Z, e) \in \text{Lip}_L(e)$ be a Lipschitz function in the exposure $e$, with Lipschitz constant $L$. Then we can bound the bias in $\widehat\tau_{DIM}$ by a multiple of $\mathcal{H}(\mathcal{C})$, and this bound is tight in a minimax sense.
\begin{lemma}\label{lem:lipschitz}
Let observations $Y_i$ be observed from a $K$-cluster design, with potential outcomes given by $Y_i(Z, e) \in \text{Lip}_L(e)$. Then the bias of the difference-in-means estimator $\widehat\tau_{DIM}$ is bounded above by
\begin{align*}
    \left\vert\E[\widehat\tau_{DIM}] - \tau^* \right\vert 
    &\leq 
    \frac{2}{N} \frac{K}{K-1} L \cdot \mathcal{H}(\mathcal{C}).
\end{align*}
Furthermore, this bound is tight over the class of Lipshitz functions,
so that balanced clustering according to the objective $\mathcal{H}(\mathcal{C})$ is tight in a minimax sense among all balanced clusterings:
\begin{align*}
    \arg\min_\mathcal{C} \max_{f\in \text{Lip}_L(e)} \left\vert\E[\widehat\tau_{DIM}] - \tau^* \right\vert
    &= \arg\min_\mathcal{C} \mathcal{H}(\mathcal{C}).
\end{align*}
\end{lemma}
We note that the objective $\mathcal{H}(\mathcal{C})$ is minimax optimal over the class of $L$-Lipschitz functions regardless of the Lipschitz constant $L$. As a result, the practitioner need not know this constant in order to find the minimax optimal clustering.

\subsection{Functions constant in neighborhoods of $\{-1, 1\}$}
A common assumption in the exposure mapping literature is that units with exposure $e_i$ close enough to their treatment assignment $Z_i$ behave as if their entire neighborhood were assigned to $Z_i$. This assumption motivates the use of IPS estimators, which are unbiased in that setting. If we let $\Delta$ denote the neighborhood of $Z_i$ in which an exposure is considered fully treated or controlled, then we have the following constraint on the potential outcome function.
\begin{align}
    \vert Y_i(Z, e) - Y_i(Z, Z) \vert &\begin{cases}
    = 0 \qquad \text{if}~ \vert Z - e\vert < \Delta\\
    \leq B \qquad \text{otherwise}
    \end{cases} \qquad \forall Z\in\{-1, 1\},~ \forall e\in[-1,1]\label{eqn:exposureMappingInterference}
\end{align}
Under this assumption on the behavior of $Y_i$, we can upper bound the bias of the difference-in-means estimator by a quantity that will turn out to be minimized by minimizing $\mathcal{H}(\mathcal{C})$.
\begin{lemma}\label{lem:exposureMapping}
Let observations $Y_i$ be observed from a $K$-cluster design, with potential outcomes satisfying Equation \eqref{eqn:exposureMappingInterference}. Then the bias of the difference-in-means estimate $\widehat\tau_{DIM}$ is bounded above by
\begin{align*}
\vert \E[\widehat\tau_{DIM}] - \tau^* \vert 
    &\leq \frac{2B}{N\Delta} \frac{K}{K-1} \mathcal{H}(\mathcal{C}).
\end{align*}
\end{lemma}
Lemma \ref{lem:exposureMapping} presents only an upper bound on the bias, and in general this bound is not tight. However, we provide simulations in Section \ref{sec:experiments:nonlinearity} showing that the objective $\mathcal{H}(\mathcal{C})$ is a reasonable heuristic for functions that satisfy Equation~\eqref{eqn:exposureMappingInterference}.

\section{Experiments}\label{sec:experiments}
We explore the performance of our cluster-based randomized design in several settings using simulated 
graphs. We compare to the baseline of unit-level randomization as well as cluster-level randomization according to several clustering schemes: the true clustering, 
direct clustering on the original bipartite graph (Section \ref{sec:design:existing}),  maximizing the $\text{Tr}(\text{Var}(\mathbf{d}))$ objective (Section \ref{sec:design:existing}), and maximizing the expected empirical variance of the doses as motivated by \cite{pouget2019variance, harshaw2022design}. 
We use code provided by the authors of \cite{aydin2019distributed} to identify a minimum-cost balanced partitioning of the graphs induced by the objectives $\mathcal{H}(\mathcal{C})$, $\text{Tr}(\text{Var}(\mathbf{d}))$, and the direct clustering.
For the last objective, which is not a balanced partitioning, we use code provided by the authors of \cite{harshaw2022design} to minimize the EXPOSURE-DESIGN objective, reporting results with the hyperparameter $\phi$ tuned to minimize mean squared error. 
See Appendix \ref{sec:algorithm} for an overview of the algorithms.

\subsection{Robustness to clusterability in the stochastic block model}\label{sec:experiments:robustness-to-graph}
We begin by studying the performance of our design as the amount of interference varies. We construct a synthetic graph according to the bipartite stochastic block model with $N=1,000$ experimental units and $M=2,000$ interference units. Both sides of the graph are partitioned into 20 equally sized groups with label $i\in[20]$. Experimental and interference units with the same label have an edge of weight $1$ with probability $0.5$, while units with different labels have an edge of weight $1$ with probability $p$. We experiment with values from $p=0$ (no interference between clusters) to $p=0.5$ (the absence of an underlying clustering structure). Potential outcomes were drawn according to the linear model \eqref{eqn:linearPotentialOutcomes}, with coefficients drawn $\alpha_i \sim \mathcal{N}(0,1)$, $\beta_i \sim \mathcal{N}(1,1)$, and $\gamma_i \sim \mathcal{N}(-1,1)$. All clustering designs used $K=20$ clusters, with $K_T=10$ clusters assigned to treatment. Table \ref{tab:robustness-to-graph} shows the relative bias (defined as $\vert\E[\widehat\tau]-\tau\vert/\tau$) of $\widehat\tau_{DIM}$ under each cluster-randomized design; variance is inconsequential in this setting, so the variance and MSE are reported in the supplementary materials. Uncertainty represents the 95\% confidence interval as determined by bootstrapping over 100 random draws of treatment assignment $\mathbf{Z}$, over a single draw of the graph and potential outcomes.

\begin{table}[]
    \centering
    \caption{Relative bias of $\widehat\tau_{DIM}$ as the bipartite stochastic block model changes (see \ref{sec:experiments:robustness-to-graph})\newline}
    \label{tab:robustness-to-graph}
    \begin{tabular}{rrrrr} 
    \toprule
    & $p=0.0$ & $p=0.005$ & $p=0.05$ & $p=0.5$\\ 
    \midrule
    $\mathcal{H}(\mathcal{C})$&\textbf{0.02($\pm$0.06)}&\textbf{3.90($\pm$0.06)}&\textbf{11.54($\pm$0.05)}&12.98($\pm$0.08)\\
    Tr(Var(d)) objective&\textbf{0.01($\pm$0.06)}&\textbf{3.84($\pm$0.05)}&\textbf{11.49($\pm$0.06)}&12.91($\pm$0.07)\\
    Direct clustering&\textbf{0.01($\pm$0.05)}&9.10($\pm$0.13)&12.68($\pm$0.07)&12.95($\pm$0.07)\\
    EXPOSURE-DESIGN &0.33($\pm$0.06)&4.06($\pm$0.06)&11.90($\pm$0.07)&13.00($\pm$0.08)\\
    Unit-level randomization&12.44($\pm$0.08)&12.55($\pm$0.08)&12.76($\pm$0.08)&12.95($\pm$0.08)\\
    \midrule
    True clusters&0.01($\pm$0.06)&3.88($\pm$0.06)&11.58($\pm$0.06)&12.96($\pm$0.06)\\
    \bottomrule
    \end{tabular}
\end{table}

As anticipated, all designs incur lower bias when there is less interference in the graph (i.e., when $p$ is smaller). Our clustering objective $\mathcal{H}(\mathcal{C})$ and the $\text{Tr}(\text{Var}(\mathbf{d}))$ objective perform on par with the true clustering for all values of $p$. The low bias of the $\mathcal{H}(\mathcal{C})$ objective is unsurprising given the minimax optimality result in Lemma \ref{lem:dim-bias}. The equivalent performance of the $\text{Tr}(\text{Var}(\mathbf{d}))$ objective is due to the symmetry inherent in the bipartite stochastic block model, in which each interference unit has the same incoming edge weight $\sum_i w_{is}$ in expectation, and the same is true of the experimental units' edge weights $\sum_s w_{is}$. In such a symmetric setting the objectives $\mathcal{H}(\mathcal{C})$ and $\text{Tr}(\text{Var}(\mathbf{d}))$ are equivalent, as can be seen by comparing their representations in Lemma \ref{lem:dim-bias} and Equation \eqref{eqn:TrVarD}.

\subsection{Robustness to nonlinearity}\label{sec:experiments:nonlinearity}
Next we explore the robustness of our design to nonlinearity in the potential outcomes model $Y_i$ by simulating outcomes according to Equation \eqref{eqn:exposureMappingInterference}, in which the potential outcome is constant for $e_i$ in a $\Delta$ neighborhood of $Z_i$. We simulate $Y_i=-Z_i$ if $|e_i - Z_i| < \Delta$ and $Y_i\sim \mathcal{U}(-1,1)$ otherwise, encoding a setting in which no knowledge can be gained about $\tau$ when $e$ is $\Delta$-far from $Z$. The graph is given by the bipartite stochastic block model described in Section \ref{sec:experiments:robustness-to-graph} with $20$ groups, with $p=0.5$ chance of an edge between units belonging to the same group and $p=0.005$ for units belonging to different groups.
All clustering designs used $K=20$ clusters, with $K_T=10$ clusters assigned to treatment.
Table \ref{tab:robustness-to-linearity} shows the results of these experiments. Uncertainty represents the 95\% confidence interval as determined by bootstrapping over 100 random draws of treatment assignment $\mathbf{Z}$, over a single draw of the graph and potential outcomes.
\begin{table}[]
    \caption{Relative bias of $\widehat\tau_{DIM}$ as the neighborhood of pure exposure, $\Delta$, widens (see \ref{sec:experiments:nonlinearity})\newline} 
    \label{tab:robustness-to-linearity}
    \centering
    \begin{tabular}{rccc} 
    \toprule
    & $\Delta= 0.1$ & $\Delta=0.3$ & $\Delta=0.5$\\
    \midrule
    $\mathcal{H}(\mathcal{C})$ &1.000($\pm$0.004)&\textbf{0.457($\pm$0.005)}&\textbf{0.001($\pm$0.000)}\\
    Tr(Var(d)) objective&1.002($\pm$0.004)&\textbf{0.460($\pm$0.004)}&\textbf{0.000($\pm$0.000)}\\
    Direct clustering&0.997($\pm$0.005)&0.950($\pm$0.008)&0.600($\pm$0.020)\\
    EXPOSURE-DESIGN &1.001($\pm$0.004)&0.509($\pm$0.005)&0.009($\pm$0.001)\\
    Unit-level randomization&0.998($\pm$0.004)&1.000($\pm$0.004)&0.998($\pm$0.003)\\
    \midrule
    True clusters&1.001($\pm$0.004)&0.458($\pm$0.004)&0.001($\pm$0.000)\\
    \bottomrule
    \end{tabular}
\end{table}

We see that all designs have lower bias when $\Delta$ is large, reflecting the fact that observations with exposure $e_i$ $\Delta$-far from $Z_i$ are useless in determining $\tau$, and it is easier to get $\Delta$-pure observations for large $\Delta$. The results in Table \ref{tab:robustness-to-linearity} support our claim that $\mathcal{H}(\mathcal{C})$ is a reasonable design heuristic for this setting (motivated by the bias upper bound in Lemma \ref{lem:exposureMapping}). We observe again that the $\text{Tr}(\text{Var}(\mathbf{d}))$ objective does as well as the $\mathcal{H}(\mathcal{C})$ objective; as described in Section \ref{sec:experiments:robustness-to-graph}, this is due to the fact that these objectives are nearly equivalent under the symmetry of our simulated graphs.

\subsection{Performance on power-law graphs}\label{sec:experiments-power-law}

To contrast with the bipartite stochastic block models studied above, we experiment on a bipartite graph model with a power-law distribution of the vertices. Our graph model combines the bipartite preferential attachment model of \cite{guillaume2006bipartite} with the affinity model of \cite{lee2015preferential}. In this model, each experimental unit is assigned to one of $K$ latent classes. For each experimental unit $i$, a degree $d_i$ is drawn from the power-law distribution $d_i=2X~:~X\sim\text{Zipf}(3)$. For each edge $j\in[d_i]$, it is either attached to an existing interference unit with probability $1-\lambda$, or to a newly drawn interference unit with probability $\lambda$. If attached an existing unit, unit $s$ is chosen with probability proportional to $(d_s + p)$ if $s$ is of the same latent class as $i$, or $d_s + q$ if $s$ is of a different class. If a new unit is drawn, then it is of the same class as unit $i$ with probability $p(p+(K-1)q)^{-1}$, and of each other class with probability $q(p+(K-1)q)^{-1}$. We chose $N=100$, $K=10$, $\lambda=0.5$, $q=0.02p$. We experimented with both $p=0.1$ (weak latent structure) and $p=100$ (strong latent structure).

Table \ref{tab:power-law} shows the relative bias of four estimators when the potential outcomes are drawn from the linear model given in Section \ref{tab:robustness-to-graph}.
Interestingly, in this setting we see that the cluster-randomized designs outperform the true latent clusters; this is possible because the interference occurs with respect to a single draw of the random graph, which the clustering algorithm gets to see. We speculate that factors specific to the power-law graphs, notably the existence of vertices with very high degree, might cause the optimal clustering for a given draw of the graph to be very different from the optimal clustering for the graph on average. This could have negative consequences for experimental design when only a random draw of the edges is observed, but the interference occurs according to the underlying latent structure. 

\begin{table}[]
    \caption{Relative bias of $\widehat\tau_{DIM}$ under an affinity model with a power-law distribution (see \ref{sec:experiments-power-law})\newline} 
    \label{tab:power-law}
    \centering
    \begin{tabular}{rccc} 
    \toprule
    & Strong latent structure & Weak latent structure\\
    \midrule
    $\mathcal{H}(\mathcal{C})$ &\textbf{1.9($\pm$0.1)}&\textbf{3.5($\pm$0.1)}\\
    Tr(Var(d)) objective&\textbf{2.0($\pm$0.1)}&3.7($\pm$0.1)\\
    Direct clustering&\textbf{1.8($\pm$0.2)}&\textbf{3.3($\pm$0.2)}\\
    Unit-level randomization&4.9($\pm$0.1)&5.4($\pm$0.1)\\
    \midrule
    True clusters&2.9($\pm$0.1)&5.2($\pm$0.1)\\
    \bottomrule
    \end{tabular}
\end{table}

\subsection{Robustness to exposure mapping: an Airbnb case study}\label{sec:experiments:airbnb}
So far we have only considered potential outcomes models where the outcome $Y_i$ is a function of the assignment $Z_i$ and the exposure $e_i$ as defined in Equation \eqref{eqn:exposureMappingDfn}. However, in many real-life settings, an exposure model is only an approximation to the true mechanism of interference. 

We test the robustness of our experimental design to misspecification of the exposure mapping by simulating outcomes according to a model developed for vacation rentals by \citet{li2022interference}, in which there is no explicit exposure mapping defined.
We call the experimental units \textit{customers} and the interference units \textit{listings}. 
In the first phase of the model, each customer $i$ applies to each listing $s$ with probability $\phi_{is}$. In the second phase, listings with applications randomly select an application to accept. 
The measured outcome $Y_i$ is $1$ if customer $i$ successfully booked a listing, and $0$ otherwise.
In alignment with previous work in this literature \cite{li2022interference, johari2022experimental} we create a natural clustering structure on the network by assigning each customer and each listing to one of 20 types.
Our simulation takes $N=500$ customers and $M=1000$ listings, with application probability under the control assignment of $\phi_{is}=0.016$ if $i$ and $s$ are of the same type, and $\phi_{is} = 0.0001$ otherwise.
Treating customer $i$ increases these application probabilities by a factor of $\alpha$.

\begin{figure}
    \centering
    \includegraphics[clip, trim=1cm 0cm 1cm 1.2cm, width=\textwidth]{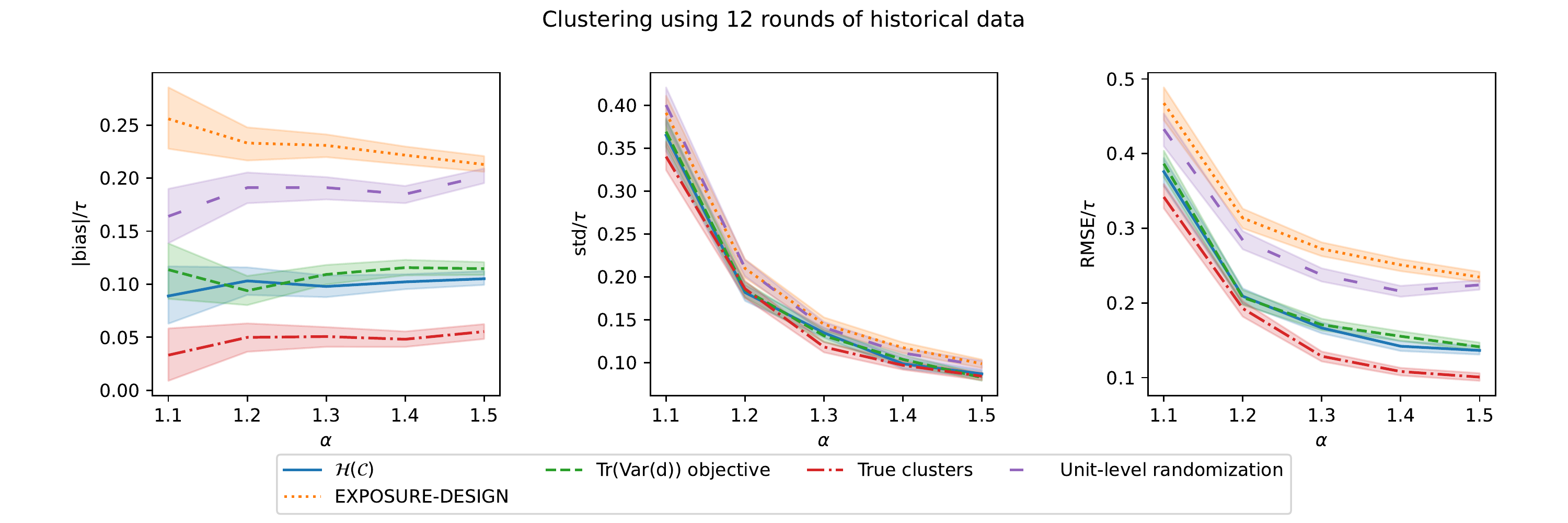}
    \caption{Performance of various clustering algorithms as the treatment effect $\alpha$ increases. Uncertainty represents 95\% empirical confidence intervals over 500 draws of the treatment assignment $\mathbf{Z}$. Direct balanced partitioning had error 4-8 times the second-worst algorithm, and is not shown.}
    \label{fig:airbnb}
\end{figure}

When running this marketplace experiment, an experimenter would typically have access to historical data about the rate of successful applications of customer $i$ to listing $s$, but would not know the consideration probabilities $\phi_{is}$. To be faithful to this observation model, we constructed the bipartite graph using twelve rounds of interaction in this marketplace under the control condition. This graph was sampled once and fixed for all experiments. All clustering designs used $K=20$ clusters, with $K_T=10$ assigned to treatment. Simulations were performed by extending code provided by the authors of \citet{li2022interference} to the cluster-randomized setting. The relative bias, standard deviation, and root mean squared error (RMSE) of $\widehat\tau_{DIM}$ of various clustering algorithms are shown in Figure \ref{fig:airbnb}.

We observe that all clustering algorithms had similar variance but that the bias of $\mathcal{H}(\mathcal{C})$ and $\text{Tr}(\text{Var}(\mathbf{d}))$ were closest to that of the true clustering, giving those two methods the lowest RMSE. Notably, the unit-randomized design is suggested by \cite{li2022interference} in this setting ($N < M$), but our cluster-randomized design outperforms it. We conclude that our clustering method still outperforms other baselines, even in a setting where the potential outcome does not obey the exposure mapping model we hypothesized.

\section*{Acknowledgements}
The authors would like to thank Christopher Harshaw, Khashayar Khosravi, Kay Brodersen, Vahan Nanumyan, and Kevin Jamieson for helpful discussions; Hannah Li and Geng Zhao for sharing their code for the Airbnb simulator; David Eisenstat for assistance with the clustering code; and Robbie Weber for feedback on the paper organization.

\newpage
\printbibliography
\newpage
\appendix

\section{Discussion of alternative normalizations of the exposure and dose models}\label{appendix:normalization}
Recall that, under the definition of exposure given in Equation \eqref{eqn:exposureMappingDfn}, both the dose $d_s$ and the exposure $e_i$ are normalized by the sum of adjacent edges so that they lie in the range $[-1,1]$. Depending on the nature of the experiment, it may instead be appropriate to define an exposure mapping in which either or both of these quantities are unnormalized, so that $\tilde{d}_s = \sum_{i} w_{is} Z_i$ and/or $\tilde{e}_i = \sum_s w_{is} d_s$.
For example, suppose that the edge weights $w_{is}$ measure the value of goods purchased by buyer $i$ from seller $s$ and that the treatment results in a multiplicative increase in the cost per unit good. 
In this setting, a more appropriate model might be one with unnormalized exposure, for example $Y_i = \gamma_i \sum_s w_{is} d_s$ where $\gamma_i$ represents the multiplicative increase in the cost per unit good for customer $i$ under treatment (the individual treatment effect).
This is the approach to normalization taken by \citet{rolnick2019randomized}, who study the bipartite model in the context of search queries issued by users (experimental units) across various geographical regions (interference units). Their potential outcomes model assumes a normalized dose but unnormalized exposure.

We note that we could also write this unnormalized exposure model in terms of the original linear model \eqref{eqn:linearPotentialOutcomes} by absorbing the exposure normalization term $(\sum_s w_{is})^{-1}$ into $\gamma_i$. However, if the normalization terms varied significantly between buyers $i$ while the individual treatment effects were approximately equal, then the minimax guarantee in Lemma \ref{lem:dim-bias} would be less meaningful when applied to the normalized model than the unnormalized exposure model.
In general, the normalization of the dose and exposure should be chosen so that the magnitude of $\gamma_i$ in the linear model \eqref{eqn:linearPotentialOutcomes} has the least variance between experimental units $i$ as possible, as this is the setting in which the minimax result of Lemma \ref{lem:dim-bias} is the tightest.

We can define the generalized objective parameterized by the definitions of dose and exposure (normalized or unnormalized) using the notation of Equation \eqref{eq:graphicalInterpretation}:
\begin{align*}
    \mathcal{H}_{e, d}(\mathcal{C})
    &= \sum_{i\in [N]}\sum_{j\not\in\mathcal{C}(i)} \gamma_i \sum_s \tfrac{1}{2}\left(e_i(\mathbf{d}_{s^+}) - e_i(\mathbf{d}_{s^-})\right)\cdot 
    \tfrac{1}{2}\left(d_s(\mathbf{Z}_{j^+}) -  d_s(\mathbf{Z}_{j^-})\right)
\end{align*}
where $\mathbf{d}_{s^+}$ (resp. $\mathbf{d}_{s^-}$) is the vector $\mathbf{d}$ with entry $d_s$ set to 1 (resp. $-1$), and analogously for $\mathbf{Z}_{j^+}$ and $\mathbf{Z}_{j^-}$.
In the setting of normalized dose and response, this simplifies to the original objective $\mathcal{H}(\mathcal{C})$. Choosing the unnormalized dose $\tilde{d}$ yields the difference $\tilde{d}_s(\mathbf{Z}_{j^+}) -  \tilde{d}_s(\mathbf{Z}_{j^-}) = 2w_{is}$, and choosing the unnormalized exposure $\tilde{e}$ yields $\tilde{e}_i(\mathbf{d}_{s^+}) - \tilde{e}_i(\mathbf{d}_{s^-}) = 2w_{js}$.

Under the linear potential outcomes model, the generalized objective satisfies the same properties as the original objective function: in particular, it is bias-minimizing in a minimax sense over all $\gamma_i\in[\Gamma_0, \Gamma_1]$, it minimizes $\text{Tr}(\text{Cov}(\mathbf{Z}, \mathbf{e}))$ for the given definition of $e$, and it has a graphical interpretation as minimizing the cut edges among an induced graph on experimental units. The first two properties can be verified via straightforward modifications of the proofs of Lemmas \ref{lem:dim-bias} and \ref{lem:covarianceObjective}, respectively, while the last property can be seen from the analogue of Equation \eqref{eq:graphicalInterpretation} in the unnormalized setting.

We conclude by discussing the relationship between the $\text{Tr}(\text{Cov}(\mathbf{Z}, \mathbf{e}))$ objective (ours) and the $\text{Tr}(\text{Var}(\mathbf{d}))$ objective (of \cite{pouget2019variance, harshaw2022design}, discussed in Section \ref{sec:design:existing}), which turn out to differ only in the normalization terms. 
We first observe that the $\text{Tr}(\text{Var}(\mathbf{d}))$ objective can be written in a form similar to that of $\mathcal{H}(\mathcal{C})$, but with a different normalization on $w_{is}$:
\begin{align}
    \arg\max_{\mathcal{C}} \text{Tr}\left( \text{Var}_{\mathbf{Z} \sim \mathcal{D}(\mathcal{C})}(\mathbf{d}) \right)
    &= \arg\min_{\mathcal{C}} \sum_{i\in[N]} \sum_{j\not\in\mathcal{C}(i)} \sum_{s\in[M]} \frac{w_{is}}{\sum_k w_{ks}} \frac{w_{js}}{\sum_k w_{ks}}. \label{eqn:TrVarD}
\end{align}
In the special case that the exposure normalization term $\sum_s w_{is}$ is equal to the dose normalization term $\sum_k w_{ks}$ for all $i$ and $s$, these constants drop out of the $\arg\min$ and the $\text{Tr}(\text{Var}(\mathbf{d}))$ objective becomes equivalent to the $\mathcal{H}(\mathcal{C})$ objective. This is approximately the setting of our synthetic experiments, in which all experimental units had the same expected number of connections (and the same for the interference units), which helps to explain why the $\text{Tr}(\text{Var}(\mathbf{d}))$ was so competitive in our experiments.    
\section{High variance of the IPS estimator}\label{appendix:ips}

In this section we substantiate the claim from Section \ref{sec:estimators} that the inverse propensity score (IPS) estimator is too high-variance to be practical in the settings we consider.
The core idea is that, when we run one-sided bipartite experiments, the definition of ``complete exposure'' to treatment or control is defined based on a unit's two-hop neighborhood. In a graph with even moderate connectivity, these neighborhoods are so large that units have a very low chance of being fully exposed to treatment or control. These small probabilities of full exposure in turn lead to high variance of the IPS estimator.

We begin by observing that, using cluster-randomized designs as we have defined them in \ref{def:balancedDesign}, there is essentially zero chance of each unit being connected completely to treatment or control under even moderate connectivity. To see this, we will compute the chance of a given experimental unit $i$ being a two-hop neighbor of unit $j$ under the bipartite stochastic block model discussed in Section \ref{sec:experiments:robustness-to-graph}. Recall that in this model, $N$ experimental units and $M$ interference units are each partitioned into $K$ equal-sized clusters labeled $1,\ldots,K$. Units with the same label have an edge between them with probability $q=0.5$, while units with different labels have an edge with probability $p$. 
We can compute the chance of experimental unit $i$ \textbf{not} having a two-hop neighbor in $\mathcal{C}(j)$ if $\mathcal{C}(i)\neq\mathcal{C}(j)$ is given by
\begin{align*}
    \P(\text{$i$ has no 2-hop neighbor in}~&\mathcal{C}(j) \vert \mathcal{C}(i)\neq\mathcal{C}(j))
    \\
    &=\prod_{k\in\mathcal{C}(j)}\prod_{s\in\mathcal{C}(j)} \P(\text{$k$ not connected to $s$ OR $i$ not connected to $s$}) \cdot\\
    &\qquad\prod_{k\in\mathcal{C}(j)}\prod_{s\in\mathcal{C}(i)} \P(\text{$k$ not connected to $s$ OR $i$ not connected to $s$})\\
    &= \prod_{k\in\mathcal{C}(j)}\prod_{s\in\mathcal{C}(j)} (1-pq) \cdot\prod_{k\in\mathcal{C}(j)}\prod_{s\in\mathcal{C}(i)} (1-pq)\\
    &= (1-pq)^{2MN/K^2}
\end{align*}
The number of clusters to which unit $i$ is connected, besides $\mathcal{C}(i)$, is the binomial random variable 
\[\text{Binom}\left(K-1, 1 - (1-pq)^{2MN/K^2}\right).\]
Observe that the probability of connection grows very quickly. In the setting of our experiments in Section \ref{sec:experiments:robustness-to-graph}, we have $M=2000$, $N=1000$, $q=0.5$ and $K=20$, so that when $p=0.005$ we have $\P(\text{$i$ has a two-hop neighbor in $\mathcal{C}(j)$} \vert \mathcal{C}(i)\neq\mathcal{C}(j)) = 1 - 1\cdot 10^{-11}$, i.e. each unit is almost certain to have a two-hop neighbor in every other cluster. We conclude that, in the graph settings we study, experimental units are likely to be connected to units of all clusters, making pure exposure to treatment or control an exceedingly rare occurrence. 

To further illustrate this point, we compute the variance of the IPS estimator on our simulated graphs for various definitions of ``pure exposure''. Under Bernoulli cluster randomization (which is slightly different from the design studied in the rest of the paper, but which is standard for the IPS estimator) we have:
\begin{align}
    \text{Var}(\widehat\tau_{IPS})
    &= \frac{1}{N^2} \sum_{i\in[N]}\frac{1}{\P(\text{unit $i$ is treated})}Y_{i,T}^2 + \frac{1}{\P(\text{unit $i$ is controlled})}Y_{i,C}^2
\end{align}
where $Y_{i,T}$ is the value of $Y_i$ under full treatment, and $Y_{i,C}$ is the value of $Y_i$ under full control. We compute the probabilities of full treatment and control via Monte Carlo simulation over draws of $\mathbf{Z}$ from a Bernoulli randomized design.

We compute the variance of the IPS estimator in two settings studied in Section \ref{sec:experiments}.

Table \ref{tab:dim-vs-ips-exposure} compares the bias of $\widehat\tau_{DIM}$ to the standard deviation of $\widehat\tau_{IPS}$ in the setting of Section \ref{sec:experiments:nonlinearity}, in which units behave as if they were fully exposed whenever $|Z_i - e_i| <\Delta$. This is precisely the setting in which the $\widehat\tau_{IPS}$ estimator works best, and it is unbiased in this case. We see from the table that the IPS estimator has less error than $\widehat\tau_{DIM}$ for small values of $\Delta$, but that when $\Delta=0.5$, the bias of $\widehat\tau_{DIM}$ has decreased to be lower than the IPS variance. We note that these figures were generated for the graph with $p=0.005$; increasing the connectivity (say, to $p=0.05$) further increases the error of $\widehat\tau_{IPS}$ relative to $\widehat\tau_{DIM}$.


\begin{table}[]
    \caption{RMSE (relative to $\tau$) of $\widehat\tau_{DIM}$ and $\widehat\tau_{IPS}$ as the neighborhood of pure exposure, $\Delta$, widens (see \ref{sec:experiments:nonlinearity})\newline} 
    \label{tab:dim-vs-ips-exposure}
    \centering
    \begin{tabular}{lccc} 
    \toprule
    & $\Delta= 0.1$ & $\Delta=0.3$ & $\Delta=0.5$\\
    \midrule
    $\widehat\tau_{DIM}$& 1.001& 0.459 & 0.001 \\
    $\widehat\tau_{IPS}$& 0.429 & 0.043 & 0.032\\
    \bottomrule
    \end{tabular}
\end{table}


Table \ref{tab:dim-vs-ips-linear} compares $\widehat\tau_{IPS}$ and $\widehat\tau_{DIM}$ in the linear setting of Section \ref{sec:experiments:robustness-to-graph}. In this setting, the IPS estimator relies on the incorrect assumption that units in a $\Delta$ neighborhood of pure exposure act as if they were purely exposed to treatment or control; this is untrue in the linear model, where even slight exposure to the opposite treatment results in a change in $Y_i$. As a result, the standard deviation reported in Table \ref{tab:dim-vs-ips-linear} provides a lower bound on the RMSE of $\widehat\tau_{IPS}$, with the remainder of the error due to bias. We see that the IPS estimator with $\Delta=0.1$ and $\Delta=0.3$ has relative error that is much higher than $\widehat\tau_{DIM}$, even before including the bias of $\widehat\tau_{IPS}$. When $\Delta=0.5$, we expect the bias of $\widehat\tau_{IPS}$ to be substantial.

\begin{table}[]
    \centering
    \caption{Bias (relative to $\tau$) of $\widehat\tau_{DIM}$ and standard deviation (relative to $\tau$) of $\widehat\tau_{IPS}$ as the bipartite stochastic block model changes (see \ref{sec:experiments:robustness-to-graph})\newline}
    \label{tab:dim-vs-ips-linear}
    \begin{tabular}{lrrrr} 
    \toprule
    & $p=0.0$ & $p=0.005$ & $p=0.05$ & $p=0.5$\\ 
    \midrule
     $\widehat\tau_{DIM}$&0.01&3.88&11.58&12.96\\
     $\widehat\tau_{IPS} (\Delta=0.1)$&0.70&9.88&390.63&508.78\\
     $\widehat\tau_{IPS} (\Delta=0.3)$&0.70&9.5&13.90&16.54\\
     $\widehat\tau_{IPS} (\Delta=0.5)$&0.70&0.71&3.70&4.31\\
    \bottomrule
    \end{tabular}
\end{table}

We conclude that, even though $\widehat\tau_{IPS}$ provides an unbiased estimate of $\tau$ when the exposure mapping is correctly specified, the variance in this estimator can be significant enough to justify using the biased estimator $\widehat\tau_{DIM}$.


\section{Counterexamples for alternative clustering methods}

\begin{figure}
    \centering
    \begin{subfigure}[b]{0.45\textwidth}
        \centering
        \includegraphics[clip, trim=0.2cm 5cm 18.6cm 2cm, width=\textwidth]{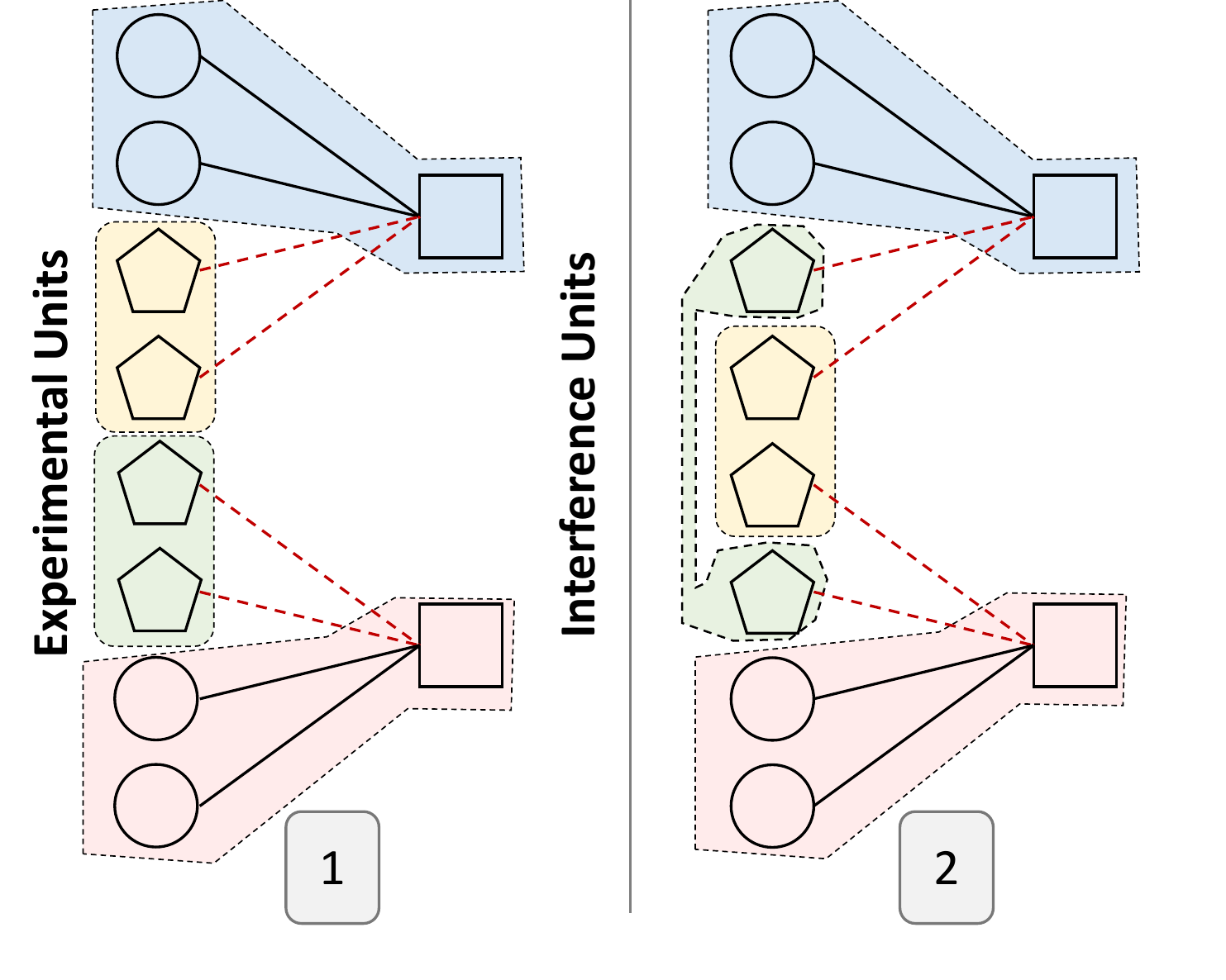}
        \caption{\textbf{Failure of the direct balanced partitioning.} 
        }
        \label{subfig:DirectClusteringCounterexample}
    \end{subfigure}
    \hfill
    \begin{subfigure}[b]{0.45\textwidth}
        \centering
        \includegraphics[clip, trim=17.8cm 5cm 1cm 2cm, width=\textwidth]{Figures/CounterExamples}
        \caption{\textbf{Failure of maximizing the dose variance.} }
        \label{subfig:VarmaxClusteringCounterexample}
    \end{subfigure}
    \caption{These counterexamples illustrate the inadequacy of existing cluster-randomized designs in the one-sided bipartite setting. In each case, Clustering 1 is the bias-minimizing clustering under our exposure model \eqref{eqn:exposureMappingDfn} and linear potential outcomes model \eqref{eqn:linearPotentialOutcomes}. 
    }
    \label{fig:counterexamples}
\end{figure}

In this section, we further expand upon the claims of section \ref{sec:design:existing} with explicit examples illustrating failures of direct balanced partitioning and of maximizing the variance among doses.
Figure \ref{fig:counterexamples} illustrates the two counterexamples.

\subsection{Direct clustering of the bipartite graph}\label{appendix:design:direct}

At a high level, the direct clustering approach fails to enforce clustering of the neighbors of interference unit $s$ that fall outside of cluster $\mathcal{C}(s)$.
This mode of failure is illustrated in Figure \ref{subfig:DirectClusteringCounterexample}, where a direct clustering of the bipartite graph assigns the same objective function score to the two clusterings, even though the first is bias-optimal under our linear model while the second is not.
Both clusterings have cost 4 according to direct partitioning (red dashed lines).
We can use the linearity of $\widehat\tau_{DIM}$ to express the bias as a sum of bias contributions of each experimental unit (see Lemma \ref{lem:unitLevelBiasDecomp}).
The bias contributed by the circular units is the same in both clusterings, since the expected exposure impurity $|e_i-Z_i|$ is the same in both clusterings and our interference model is linear in $e_i$ and $Z_i$.
However, the exposure of the pentagonal units in Clustering 1 stochastically dominates that of Clustering 2 in the sense of being closer to the true treatment assignment $Z_i$, which means that Clustering 1 incurs less bias. 
Therefore, Clustering 1 is superior under our potential outcomes model.

\subsection{Maximizing the variance of the doses}\label{appendix:design:varmax}

Figure \ref{subfig:VarmaxClusteringCounterexample} exhibits a bipartite graph in which the doses $d_s$ are primarily controlled by a small number of experimental units with highly weighted edges, so that maximizing $\text{Tr}(\text{Var}(\mathbf{d}))$ actually reduces the average covariance of $Z_i$ and $e_i$.
Clustering 1 ensures that the majority of experimental units (the $2n$ pentagons) have $e_i=Z_i$ w.p. $\approx\tfrac{1}{2}$, and $e_i = \tfrac{3}{4}Z_i$ otherwise, while sacrificing the exposure of the two middle circular units, which have $e_i=0$ w.p. $\approx\tfrac{1}{2}$. By contrast, Clustering 2 obtains a high dose variance by prioritizing the assignment of the middle circular units to the same cluster, at the expense of making $e_i=0$ w.p. $\tfrac{1}{2}$ for all $2n$ pentagonal units. If we think of the interference among all units as being on the same order of magnitude then Clustering 1, which incurs less interference in the $2n$ pentagonal units, is the correct choice as $n\to\infty$. If, on the other hand, the potential outcomes scale in magnitude with the sum of an experimental unit's edge weights, then there are settings in which Clustering 2 is bias-optimal. This can be addressed by choosing the appropriate normalization of the dose and exposure for the problem at hand and modifying $\mathcal{H}(\mathcal{C})$ to reflect the choice of normalization; see Appendix \ref{appendix:normalization} for further discussion.
\section{Comparison to two-sided designs}
A recent development in the bipartite design literature is the two-sided randomization designs of \citet{bajari2021multiple} and \citet{johari2022experimental}, in which experimental and interference units are each randomized to treatment and control, but treatment is only applied when a treated unit interacts with another treated unit. The major benefit of this design is that it allows for the quantification of spillover effects by comparing the fully controlled interactions (in which both sides of the interaction were controlled) to the interactions that experience interference (in which only one side of the interaction was assigned to control). In the absence of interference, these interaction types should have the same average response; any deviation from equality indicates interference that can be quantified.

The idea of measuring and correcting for interference is very appealing, especially when interference can be quantified using the same experiment that measures the global treatment effect. We think that this line of work is an important part of mitigating interference, and that the two-sided randomization design is a particularly clever way of approaching the problem. One challenge with TSR designs, as discussed by \citet[Section 6]{bajari2021multiple}, is that without further assumptions on the interference model, TSR with treatment fraction $p$ is only able to estimate the global effect of treating $p$-fraction of the units as compared with treating no units, instead of the global effect of treating all units as compared with treating no units. \citet[Section 8.1]{bajari2021multiple} suggests randomizing at the interaction level to allow estimation of the spillover effect at multiple levels of interaction, which could then be extrapolated to estimate the global treatment effect - this of course relies on some model for the extrapolation. Ultimately, then, one difference between the TSR design and the cluster-randomized design is that the latter tries to get as many units in ``pure treatment'' and ``pure control'' as possible, to estimate the global treatment effect while limiting the need for such extrapolation. As \citet[Section 8]{johari2022experimental} suggests, cluster-randomization is likely to outperform TSR when the underlying graph is well-clusterable, but TSR is likely to do better when no cluster-based method can achieve near-complete treatment or control of units.

It would be interesting to think about combinations of TSR and cluster-based designs, particularly in the context of the extrapolation ideas proposed in \citet[Section 8.1]{bajari2021multiple}. One option would be to cluster-randomize the buyers and sellers individually, in a way that maximizes the variance of the unit-level exposures to treatment under a TSR design (echoing the objective of \citet{pouget2019variance} and \citet{harshaw2022design}). The TSR design would allow for the estimation of spillover effects at various levels of interference, while maximizing the variance among the realized levels of interference would improve the accuracy of the extrapolation. 

\section{Considerations for balanced clustering}\label{sec:future}
In this appendix, we discuss potential benefits and detriments of cluster-randomized designs, expanding upon discussion in Section \ref{sec:experimentalDesign}.

\subsection{Implementation benefits of balanced clusters}
Practitioners may appreciate two benefits of balanced designs beyond the variance reduction mentioned in Section \ref{sec:experimentalDesign}. First, when exactly $K_T$ of $K$ clusters are treated, balancing the clusters ensures control over the fraction of units that are treated. Controlling this fraction is important when we want to balance the scientific value of experimentation with potential negative effects on the treated units (i.e. staying within the experimental budget). Secondly, it has been our experience that many clustering algorithms that do not control for balancedness and cardinality sometimes produce many singletons clusters, when clustering incentives are not strong enough to group these units with other units. This is what we observed with the Exposure-Design objective of \citet{harshaw2022design} when its hyperparameters are not tuned properly. When datasets are quite large, the large number of singleton clusters produced by these algorithms can slow down certain data analysis pipelines that work better in low cardinality settings. If these singletons clusters are to be clustered to reduce cardinality without improving any ``cut''-like objective, it may make sense to do so in a balanced way for any of the reasons listed above. In other words, the occasional practical need to control for cluster cardinality is extra motivation to maintain balance instead of an arbitrary grouping of isolated nodes.

\subsection{Considerations in the presence of cluster-based heterogeneity}
We studied a clustering objective, $\mathcal{H}(\mathcal{C})$, that minimizes the bias of the difference-in-means estimator for one-sided bipartite experiments. When the experimental units are well-clusterable, 
the cluster-randomized design significantly reduces the bias in $\widehat\tau_{DIM}$ when compared with unit-level randomization. However, cluster-based randomization may actually reduce the accuracy of the treatment effect estimate if the clustering correlates with individual treatment effects $\tau_i := Y_i(\mathbf{Z}=\mathbf{1}) - Y_i(\mathbf{Z}=-\mathbf{1})$. This may result in significant heterogeneity in the average treatment effect among clusters, which introduces additional variance in the difference-in-means estimator. In the extreme case, the bias-variance tradeoff may favor a unit-randomized design over a cluster-randomized design.
We note that this problem is not unique to $\widehat\tau_{DIM}$; the IPS estimator also experiences a tradeoff between the variance reduction due to clustering (due to a higher probability of pure exposure) and the variance increase due to heterogeneity between clusters.
In practice the variance is typically reduced by choosing a large number of clusters, since the cluster becomes the effective unit of analysis.
Of interest is understanding the trade-off incurred by cluster randomization in this setting, and to design techniques to determine, perhaps from historical data, whether cluster randomization should be used for a given experiment.

\section{Proofs}\label{appendix:proofs}

In this appendix, we provide proofs of all the results in this paper. We apply similar proof techniques to bound the bias in a variety of potential outcomes models, and therefore consider a potential outcomes model which generalizes all the models discussed in this paper:
\begin{align*}
    Y_i &= g_i(Z_i, e_i).
\end{align*}
We will first prove some useful lemmas that apply to all potential outcomes models of this form; later we will provide results that are specific to each model type.

\subsection{Useful Lemmas}

The first two lemmas in this section will be useful for calculating the bias of the difference in means estimator under the linear, Lipschitz, and $\Delta-$neighborhood functions. The proofs for each bias bound will follow a similar structure: First, we will decompose the bias of the difference-in-means estimator into the bias contribution of each unit under both treatment and control. Next, we will use the given structure of the potential outcome to either compute the bias exactly (in the linear setting) or bound the bias (in the Lipschitz and $\Delta-$neighborhood settings). In each case, we will be left with a bound in terms of the conditional expectation of the exposure $e_i$ given the treatment status $Z_i$. Our third step will be to relate this quantity to the folded graph clustering objective, $\mathcal{H}(\mathcal{C})$.

We provide lemmas for the first and third steps; the second step is unique to each bias calculation.

\begin{lemma}[Unit-level bias decomposition]\label{lem:unitLevelBiasDecomp}
Let the unit-level responses $Y_i$ be functions of the unit's treatment $Z_i$ and the unit-level exposure $e_i$, so that $Y_i = g_i(Z_i, e_i)$. Then the bias of the difference-in-means estimate of the average treatment effect can be written as
{\small
\begin{align*}
    \E[\hat\tau] - \tau^*
    &= \frac{1}{N}\sum_{i\in[N]} \E[g_i(Z_i, e_i) - g_i(Z_i, Z_i) \vert Z_i = 1] - \frac{1}{N}\sum_{i\in[N]} \E[g_i(Z_i, e_i) - g_i(Z_i, Z_i) \vert Z_i = -1]
\end{align*}
}
\end{lemma}
\begin{proof}
We begin by writing the bias in terms of the response function $g_i(Z_i, e_i)$:
{\small
\begin{align*}
    \E[\hat\tau] - \tau^* &= \E\left[ \frac{1}{N_T} \sum_{i\in I} Y_i - \frac{1}{N_C} \sum_{i\in\bar{I}} Y_i \right] - \frac{1}{N} \sum_{i\in[N]} (g_i(1, 1) - g_i(-1, -1))\\
    &=  \frac{1}{N_T} \sum_{i\in [N]} \E\left[\mathbf{1}\{Z_i=1\}Y_i\right] - \frac{1}{N_C} \sum_{i\in[N]} \E\left[\mathbf{1}\{Z_i=-1\}Y_i\right]  - \frac{1}{N} \sum_{i\in[N]} (g_i(1, 1) - g_i(-1, -1))\\
    &=  \frac{1}{N_T} \sum_{i\in [N]} \E\left[\mathbf{1}\{Z_i=1\}g_i(Z_i, e_i)\right] - \frac{1}{N_C} \sum_{i\in[N]} \E\left[\mathbf{1}\{Z_i=-1\}g_i(Z_i, e_i)\right]  \\
    &\qquad\qquad-\frac{1}{N} \sum_{i\in[N]} (g_i(1, 1) - g_i(-1, -1))
\end{align*}
}
We use the law of total probability to rewrite the expectations as conditional expectations:
\begin{align*}
    \E[\hat\tau] - \tau^* &= \frac{1}{N_T} \sum_{i\in [N]} \E\left[g_i(Z_i, e_i)\vert Z_i = 1\right] \P(Z_i = 1)\\&\qquad\qquad - \frac{1}{N_C} \sum_{i\in[N]} \E\left[g_i(Z_i, e_i) \vert Z_i = -1\right] \P(Z_i = -1)  \\&\qquad\qquad -\frac{1}{N} \sum_{i\in[N]} (g_i(1, 1) - g_i(-1, -1))\\
    &= \frac{1}{N_T} \sum_{i\in [N]} \frac{N_T}{N}\E\left[g_i(Z_i, e_i)\vert Z_i = 1\right]\\&\qquad\qquad - \frac{1}{N_C} \sum_{i\in[N]} \frac{N_C}{N}\E\left[g_i(Z_i, e_i) \vert Z_i = -1\right] \\&\qquad\qquad - \frac{1}{N} \sum_{i\in[N]} (g_i(1, 1) - g_i(-1, -1))\\
    &= \frac{1}{N} \sum_{i\in [N]} \E\left[g_i(Z_i, e_i)\vert Z_i = 1\right]\\&\qquad\qquad - \frac{1}{N} \sum_{i\in[N]} \E\left[g_i(Z_i, e_i) \vert Z_i = -1\right] \\&\qquad\qquad - \frac{1}{N} \sum_{i\in[N]} (g_i(1, 1) - g_i(-1, -1))
\end{align*}
All of the summations are now computing averages over the $N$ terms. We can distribute the terms of the final summation between the first and second summations, which lets us decompose the bias into contributions from each unit under its control and treated assignments:
{\small
\begin{align*}
    \E[\hat\tau] - \tau^* &= \frac{1}{N} \sum_{i\in [N]} \E\left[g_i(Z_i, e_i) - g_i(1,1)\vert Z_i = 1\right] - \frac{1}{N} \sum_{i\in[N]} \E\left[g_i(Z_i, e_i) - g_i(-1, -1) \vert Z_i = -1\right] \\
    &= \frac{1}{N} \sum_{i\in [N]} \E\left[g_i(Z_i, e_i) - g_i(Z_i,Z_i)\vert Z_i = 1\right] - \frac{1}{N} \sum_{i\in[N]} \E\left[g_i(Z_i, e_i) - g_i(Z_i, Z_i) \vert Z_i = -1\right]
\end{align*}
}
as desired.
\end{proof}

\begin{lemma}[Writing the weighted condition gaps in terms of the graph structure]\label{lem:conditionalSumToGraphObjective}
Let $e_i$ be the exposure of unit $i$ to treatment, as defined in Eqn \eqref{eqn:exposureMappingDfn}. Let $\gamma_i\in\R$ be arbitrary. If the treatment assignment vector $\mathbf{Z}\sim\mathcal{D}(\mathcal{C})$ is drawn according to a balanced cluster randomized design (definition \ref{def:balancedDesign}), then the  $\gamma_i$-weighted average conditional gap between $e_i$ and the unit's exposure $Z_i$ can be written in terms of the underlying graph weights $w_{is}$ between units assigned to different clusters:
{\small
\begin{align*}
    \frac{1}{N} \sum_{i\in [N]} \gamma_i \left(\E\left[Z_i - e_i \vert Z_i = 1\right] +  \E\left[e_i - Z_i \vert Z_i = -1\right]\right)
    &= \frac{2}{N} \frac{K}{K-1} \sum_{i\in[N]} \sum_{j\not\in\mathcal{C}(i)} \gamma_i  \sum_{s} \frac{w_{is}}{\sum_s w_{is}} \frac{w_{js}}{\sum_k w_{ks}}.
\end{align*}
}
\end{lemma}
\begin{proof}
We begin by observing the useful fact that
\begin{align*}
    \sum_s \frac{w_{is}}{\sum_s w_{is}} \sum_j \frac{w_{js}}{\sum_j w_{js}} &= 1.
\end{align*}
Combining this fact with the definition of $e_i$ lets us write all of the terms in this expression as linear combinations of $Z_j$ and $Z_i$, where $Z_i$ is fixed in each conditional expectation.
\begin{align*}
    \frac{1}{N} \sum_{i\in [N]}& \gamma_i \left(\E\left[Z_i - e_i \vert Z_i = 1\right] +  \E\left[e_i - Z_i \vert Z_i = -1\right]\right)\\
    &= \frac{1}{N} \sum_{i\in[N]} \gamma_i \Bigg(\E\left[ \sum_s \frac{w_{is}}{\sum_s w_{is}} \sum_j \frac{w_{js}}{\sum_j w_{js}} (Z_i - Z_j) \vert Z_i=1\right] \\
    &\qquad\qquad
    + \E\left[ \sum_s \frac{w_{is}}{\sum_s w_{is}} \sum_j \frac{w_{js}}{\sum_j w_{js}} (Z_j - Z_i) \vert Z_i=-1\right] 
    \Bigg)\\
    &= \frac{1}{N} \sum_{i\in[N]} \gamma_i  \sum_s \frac{w_{is}}{\sum_s w_{is}} \sum_j \frac{w_{js}}{\sum_j w_{js}} \left(\E\left[(Z_i - Z_j) \vert Z_i=1\right] 
    + \E\left[ (Z_j - Z_i) \vert Z_i=-1\right] 
    \right)\\
    &= \frac{1}{N} \sum_{i\in[N]} \gamma_i  \sum_s \frac{w_{is}}{\sum_s w_{is}} \sum_j \frac{w_{js}}{\sum_j w_{js}} \left(2\P\left(Z_j = -1 \vert Z_i=1\right) 
    + 2\P\left( Z_j = 1 \vert Z_i=-1\right)
    \right)
\end{align*}
where in the last step we used the fact that $Z_j$ only takes values in $\{-1,1\}$. 

Recall that the $Z_i$ were assigned according to a \textit{cluster-randomized} design $\mathcal{C}$. If $j\in\mathcal{C}(i)$ then $Z_i=Z_j$, so that units in the same cluster contribute zero to the summation above. Otherwise, if $j\not\in\mathcal{C}(i)$, we can compute the probability that $Z_i \neq Z_j$. If there are $K$ clusters and $K_T$ of them are chosen to be treated, then we have
\begin{align*}
    \P(Z_i = -1 | Z_j = 1 \cap j\not\in\mathcal{C}(i)) &= 
    \frac{K_C}{K-1}
\end{align*}
and
\begin{align*}
    \P(Z_i =1 | Z_j = -1 \cap j\not\in\mathcal{C}(i)) &= 
    \frac{K_T}{K-1}.
\end{align*}
We can use this information to compute the probabilities in the expression above, applying the fact that $\P(Z_i \neq Z_j \vert j\in\mathcal{C}(i)) = 0$ to restrict the sum over $j$ to only the units outside of $\mathcal{C}(i)$.
\begin{align*}
    \frac{1}{N} \sum_{i\in [N]}& \gamma_i \left(\E\left[Z_i - e_i \vert Z_i = 1\right] +  \E\left[e_i - Z_i \vert Z_i = -1\right]\right)\\
    &= \frac{2}{N} \sum_{i\in[N]} \gamma_i  \sum_s \frac{w_{is}}{\sum_s w_{is}} \sum_{j\not\in\mathcal{C}(i)} \frac{w_{js}}{\sum_j w_{js}} \Big(\P(Z_j = -1 \vert Z_i=1\cap j\not\in\mathcal{C}(i) ) 
    \\&\qquad\qquad\qquad\qquad\qquad\qquad\qquad\qquad\qquad
    + \P( Z_j = 1 \vert Z_i=-1\cap j\not\in\mathcal{C}(i))
    \Big)\\
    &= \frac{2}{N} \sum_{i\in[N]} \gamma_i  \sum_s \frac{w_{is}}{\sum_s w_{is}} \sum_{j\not\in\mathcal{C}(i)} \frac{w_{js}}{\sum_j w_{js}} \left( \frac{K_C}{K-1}
    + \frac{K_T}{K-1}
    \right)\\
    &= \frac{2}{N} \sum_{i\in[N]} \gamma_i \sum_s \frac{w_{is}}{\sum_s w_{is}} \sum_{j\not\in\mathcal{C}(i)} \frac{w_{js}}{\sum_j w_{js}} \frac{K}{K-1},
\end{align*}
as desired.
\end{proof}

The final lemma in this section writes the exposure vector $\mathbf{e}$ is a linear combination of the treatment assignments $\mathbf{Z}$. This linearity is useful for establishing the connection between the covariance objective and the folded graph objective in Lemma \ref{lem:covarianceObjective}.

\begin{lemma}\label{lem:e=CZ}
Let $\mathbf{e}$ be defined as in Equation \eqref{eqn:exposureMappingDfn}. Then $\mathbf{e} = C \mathbf{Z}$, where 
\begin{align*}
    C_{ij} &= \sum_s \frac{w_{is}}{\sum_s w_{is}} \frac{w_{js}}{\sum_k w_{ks}}
\end{align*}
\end{lemma}
\begin{proof}
The proof proceeds by the definition of $e_i$. Let matrix $B\in[0,1]^{N\times M}$ be defined as 
\begin{align*}
    B_{is} &= \frac{w_{is}}{\sum_s w_{is}}.
\end{align*}
Then we can write the exposures $e_i$ as a linear combination of the doses $d_s$:
\begin{align*}
    e_i &= \frac{\sum_s w_{is} d_s}{\sum_s w_{is}}\\
    &= [B \mathbf{d}]_i.
\end{align*}
Similarly, let matrix $A\in[0,1]^{M\times N}$ be defined as
\begin{align*}
    A_{si} &= \frac{w_{is}}{\sum_i w_{is}},
\end{align*}
so that we can write the doses $d_s$ as a linear combination of the treatment effects $\mathbf{Z}$:
\begin{align*}
    d_s &= \frac{\sum_i w_{is} Z_i}{\sum_i w_{is}}\\
    &= [A\mathbf{Z}]_s.
\end{align*}
Putting these together, we have
\begin{align*}
    \mathbf{e} &= BA\mathbf{Z}\\
    &=: C\mathbf{Z}
\end{align*}
with $C_{ij}$ as given in the lemma statement.
\end{proof}
With these helper lemmas established, we turn to proving the main results in the paper.

\subsection{Proof of Lemma \ref{lem:dim-bias}}
We begin by applying Lemma \ref{lem:unitLevelBiasDecomp} to decompose the bias in terms of its unit-level contributions:
{\small
\begin{align*}
    \E[\hat\tau] - \tau^* &= \frac{1}{N} \sum_{i\in [N]} \E\left[g_i(Z_i, e_i) - g(Z_i,Z_i)\vert Z_i = 1\right] - \frac{1}{N} \sum_{i\in[N]} \E\left[g_i(Z_i, e_i) - g(Z_i, Z_i) \vert Z_i = -1\right]
\end{align*}
}
We apply the linear response function to simplify this expression:
\begin{align*}
    \E[\hat\tau] - \tau^* &= \frac{1}{N} \sum_{i\in [N]} \E\left[(\alpha_i + \beta_i Z_i + \gamma_i e_i) - (\alpha_i + \beta_i Z_i + \gamma_i Z_i)\vert Z_i = 1\right] \\&\qquad\qquad- \frac{1}{N} \sum_{i\in[N]} \E\left[(\alpha_i + \beta_i Z_i + \gamma_i e_i) - (\alpha_i + \beta_i Z_i + \gamma_i Z_i) \vert Z_i = -1\right]\\
    &= \frac{1}{N} \sum_{i\in [N]} \E\left[\gamma_i e_i -  \gamma_i Z_i\vert Z_i = 1\right] - \frac{1}{N} \sum_{i\in[N]} \E\left[\gamma_i e_i - \gamma_i Z_i \vert Z_i = -1\right]\\
    &= -\frac{1}{N} \sum_{i\in [N]} \gamma_i \left(\E\left[Z_i - e_i \vert Z_i = 1\right] + \E\left[e_i - (-1) \vert Z_i = -1\right]\right)
\end{align*}

Next, we apply Lemma \ref{lem:conditionalSumToGraphObjective} to write this expression in terms of the graph structure, completing the proof of the first statement of the lemma:
\begin{align}
    \E[\widehat\tau] - \tau^* 
    &= -\frac{2}{N} \frac{K}{K-1} \sum_{i\in[N]} \sum_{j\not\in\mathcal{C}(i)} \gamma_i  \sum_{s} \frac{w_{is}}{\sum_s w_{is}} \frac{w_{js}}{\sum_k w_{ks}}.\label{eqn:linear-bias}
\end{align}

Now we will prove the second part of the lemma, that the folded graph objective provides the bias-minimizing clustering in a minimax sense among all balanced cluster-randomized designs when the potential outcome is linear in $Z$ and the exposure $e$, and the interference parameter $\gamma$ satisfies $\gamma_i = O(1)$. In particular, we will prove this claim when the $\gamma_i$ are bounded on a shared interval $[\Gamma_0, \Gamma_1]$ for all $i$.

We begin by finding the maximum bias (over choice of $\mathbf{\gamma}$) for a given clustering $\mathcal{C}$. We use the value of the bias from Eqn \eqref{eqn:linear-bias}:
\begin{align*}
    \arg\min_{\mathcal{C}} \max_{\mathbf{\gamma}\in[\Gamma_0, \Gamma_1]} &\left| \E_{\mathbf{Z} \sim \mathcal{D}(\mathcal{C})}[\hat\tau] - \tau^*  \right|\notag\\
    &= \arg\min_{\mathcal{C}} \max_{\mathbf{\gamma}\in[\Gamma_0, \Gamma_1]^N} 2\frac{1}{N}\cdot\frac{K}{K-1}\left| \sum_{i\in[N]} \sum_{j\not\in\mathcal{C}(i)} \gamma_i  \sum_{s} \frac{w_{is}}{\sum_s w_{is}} \frac{w_{js}}{\sum_k w_{ks}} \right|.
\end{align*}
Recall that the edge weights $w_{is}$ are all nonnegative, so term $i$ of the summation takes on the sign of $\gamma_i$.
For this reason, the maximum bias occurs when all $\gamma_i$ are of the same sign (so that no terms in the summation cancel each other), at $\gamma_i = \max(\vert\Gamma_0\vert, \vert\Gamma_1\vert)$.
\begin{align*}
    \arg\min_{\mathcal{C}} \max_{\mathbf{\gamma}\in[\Gamma_0, \Gamma_1]} &\left| \E_{\mathbf{Z} \sim \mathcal{D}(\mathcal{C})}[\hat\tau] - \tau^*  \right|\notag\\
    &= \arg\min_{\mathcal{C}} \max(\vert\Gamma_0\vert, \vert\Gamma_1\vert) \cdot 2\frac{1}{N}\cdot\frac{K}{K-1} \sum_{i\in[N]} \sum_{j\not\in\mathcal{C}(i)}  \sum_{s} \frac{w_{is}}{\sum_s w_{is}} \frac{w_{js}}{\sum_k w_{ks}}.
\end{align*}
The terms $K$, $N$, $\Gamma_0$ and $\Gamma_1$ are all constants with respect to the clustering $\mathcal{C}$, so they can be removed without affecting the arg max. We recognize this objective as precisely our folded graph objective
\begin{align*}
    \arg\min_{\mathcal{C}} \max_{\mathbf{\gamma}\in[\Gamma_0, \Gamma_1]} \left| \E_{\mathbf{Z} \sim \mathcal{D}(\mathcal{C})}[\hat\tau] - \tau^*  \right|
    &= \arg\min_{\mathcal{C}}H(\mathcal{C})
\end{align*}
as desired.
\qed

\subsection{Proof of Lemma \ref{lem:lipschitz} (bounding the bias under the Lipschitz potential outcomes model)}

We begin by applying Lemma \ref{lem:unitLevelBiasDecomp} to decompose the bias in terms of its unit-level contributions:
{\small
\begin{align}
    \left\vert\E[\hat\tau] - \tau^* \right\vert&= \left\vert\frac{1}{N} \sum_{i\in [N]} \E\left[g_i(Z_i, e_i) - g(Z_i,Z_i)\vert Z_i = 1\right] - \frac{1}{N} \sum_{i\in[N]} \E\left[g_i(Z_i, e_i) - g(Z_i, Z_i) \vert Z_i = -1\right]\right\vert\notag\\
    &= \left\vert\frac{1}{N} \sum_{i\in [N]} \E\left[g_i(1, e_i) - g(1,Z_i)\vert Z_i = 1\right] - \frac{1}{N} \sum_{i\in[N]} \E\left[g_i(-1, e_i) - g(-1, Z_i) \vert Z_i = -1\right]\right\vert \label{eqn:Lipschitz-sum}
\end{align}
}
Next, we use the fact that $g_i(Z, e)$ is $L$-Lipschitz in $e$ to simplify this expression:
\begin{align*}
    \left\vert\E[\hat\tau] - \tau^* \right\vert &\leq
    \frac{1}{N} \sum_{i\in [N]} \E\left[L\left\vert Z_i - e_i\right\vert \big\vert Z_i = 1\right] + \frac{1}{N} \sum_{i\in[N]} \E\left[L\left\vert e_i - Z_i\right\vert \big\vert Z_i = -1\right]\\
    &=
    \frac{1}{N} \sum_{i\in [N]}L \E\left[\left\vert 1 - e_i\right\vert \big\vert Z_i = 1\right] + \frac{1}{N} \sum_{i\in[N]}L \E\left[\left\vert e_i - (-1)\right\vert \big\vert Z_i = -1\right].
\end{align*}
Observe that $e_i \in [-1, 1]$, so that the terms in absolute values are all positive. This lets us drop the absolute value signs and write
\begin{align*}
    \left\vert\E[\hat\tau] - \tau^* \right\vert &\leq
    \frac{1}{N} \sum_{i\in [N]}L \E\left[ 1 - e_i \vert Z_i = 1\right] + \frac{1}{N}  \sum_{i\in[N]}L \E\left[ e_i - (-1) \vert Z_i = -1\right].
\end{align*}
We apply Lemma \ref{lem:conditionalSumToGraphObjective} with $\gamma_i = L$ to bound the bias as
\begin{align*}
    \left\vert\E[\hat\tau] - \tau^* \right\vert &\leq
    \frac{2}{N} \frac{K}{K-1} L \sum_{i\in[N]} \sum_{j\not\in\mathcal{C}(i)} \frac{1}{\sum_s w_{is}} w_i^T \tilde{w}_j.
\end{align*}
To prove the second part of the lemma statement (minimax optimality), we compute the maximum bias over $L$-Lipschitz interference functions $g_i$, using the expression for bias given in Equation \eqref{eqn:Lipschitz-sum}.
\begin{align}
    \arg\min_\mathcal{C} \max_{\{g_i\in\text{Lip}_L(e)\}} \left\vert\E[\hat\tau] - \tau^* \right\vert
    &= \arg\min_\mathcal{C} \max_{\{g_i\in\text{Lip}_L(e)\}} \Bigg\vert\frac{1}{N} \sum_{i\in [N]} \E\left[g_i(1, 1) - g_i(1, e_i)\vert Z_i = 1\right] 
    \notag\\
    &\qquad\qquad\qquad + \frac{1}{N} \sum_{i\in[N]} \E\left[g_i(-1, e_i) - g_i(-1, -1) \vert Z_i = -1\right]\Bigg\vert.\label{eqn:lipschitzMinimaxStatement}
\end{align}
Next, we will show that setting $g_i(Z, e) = L\cdot e~\forall i$ achieves the maximum over all $L$-Lipschitz functions by computing an upper bound on the argument of the maximum and showing that this choice of $\{g_i\}$ attains that bound. We begin by upper bounding the maximum:
\begin{align*}
    \max_{\{g_i\in\text{Lip}_L(e)\}} &\left\vert\frac{1}{N} \sum_{i\in [N]} \E\left[g_i(1, 1) - g_i(1, e_i)\vert Z_i = 1\right] + \frac{1}{N} \sum_{i\in[N]} \E\left[g_i(-1, e_i) - g_i(-1, -1) \vert Z_i = -1\right]\right\vert\\
    &\leq \max_{\{g_i\in\text{Lip}_L(e)\}}\Bigg(\frac{1}{N} \sum_{i\in [N]} \E\left[\left\vert g_i(1, 1) - g_i(1, e_i)\right\vert\big\vert Z_i = 1\right] \\
    &\qquad\qquad\qquad\qquad\qquad + \frac{1}{N} \sum_{i\in[N]} \E\left[\left\vert g_i(-1, e_i) - g_i(-1, -1)\right\vert \big\vert Z_i = -1\right]\Bigg)\\
    &\leq \frac{1}{N} \sum_{i\in [N]} \E\left[\max_{g_i\in\text{Lip}_L(e)}\left\vert g_i(1, 1) - g_i(1, e_i)\right\vert\big\vert Z_i = 1\right] \\&\qquad+\frac{1}{N} \sum_{i\in[N]} \E\left[\max_{g_i\in\text{Lip}_L(e)}\left\vert g_i(-1, e_i) - g_i(-1, -1)\right\vert \big\vert Z_i = -1\right]
\end{align*}
Next we apply the Lipschitz assumption, the fact that $e_i\in[-1,1]$, and Lemma \ref{lem:conditionalSumToGraphObjective} with $\gamma_i=L$:
\begin{align*}
    \max_{\{g_i\in\text{Lip}_L(e)\}} &\left\vert\frac{1}{N} \sum_{i\in [N]} \E\left[g_i(1, 1) - g_i(1, e_i)\vert Z_i = 1\right] + \frac{1}{N} \sum_{i\in[N]} \E\left[g_i(-1, e_i) - g_i(-1, -1) \vert Z_i = -1\right]\right\vert\\
    &\leq \frac{1}{N} \sum_{i\in [N]} \E\left[L\left\vert 1 - e_i\right\vert\big\vert Z_i = 1\right] + \frac{1}{N} \sum_{i\in[N]} \E\left[L\left\vert e_i - (-1)\right\vert \big\vert Z_i = -1\right]\\
    &= \frac{1}{N} \sum_{i\in [N]} L \E\left[1 - e_i\big\vert Z_i = 1\right] + \frac{1}{N} \sum_{i\in[N]} L \E\left[ e_i - (-1) \big\vert Z_i = -1\right]\\
    &= \frac{2}{N} \frac{K}{K-1} L \sum_{i\in[N]} \sum_{j\not\in\mathcal{C}(i)} \sum_{s} \frac{w_{is}}{\sum_s w_{is}} \frac{w_{js}}{\sum_k w_{ks}}.
\end{align*}
We have now shown an upper bound on the argument of the right hand side of Equation \eqref{eqn:lipschitzMinimaxStatement}. Next, we will show that $\tilde{g}_i(Z, e) := L\cdot e ~\forall i$ achieves this bound (and is therefore a maximizer over the class of $L$-Lipschitz functions) by directly computing the argument from the right hand side of Equation \eqref{eqn:lipschitzMinimaxStatement} under this choice of $g$. We use the definition of $\tilde{g}$ to bound the difference in $\tilde{g}$ by the difference in $e$, the fact that $e_i\in[-1,1]$ to remove the absolute values, and finally apply Lemma \ref{lem:conditionalSumToGraphObjective} with $\gamma_i=L$.
\begin{align}
    \Big\vert\frac{1}{N} \sum_{i\in [N]} \E[\tilde{g}_i(1, 1) &- \tilde{g}_i(1, e_i)\vert Z_i = 1] + \frac{1}{N} \sum_{i\in[N]} \E\left[\tilde{g}_i(-1, e_i) - \tilde{g}_i(-1, -1) \vert Z_i = -1\right]\Big\vert\notag\\
    &=\left\vert\frac{1}{N} \sum_{i\in [N]} \E\left[L (1 - e_i)\vert Z_i = 1\right] + \frac{1}{N} \sum_{i\in[N]} \E\left[L(e_i - (-1)) \vert Z_i = -1\right]\right\vert\notag\\
    &=\frac{1}{N} \sum_{i\in [N]}L \E\left[ (1 - e_i)\vert Z_i = 1\right] + \frac{1}{N} \sum_{i\in[N]}L \E\left[(e_i - (-1)) \vert Z_i = -1\right]\notag\\
    &=\frac{2}{N} \frac{K}{K-1} L \sum_{i\in[N]} \sum_{j\not\in\mathcal{C}(i)} \sum_{s} \frac{w_{is}}{\sum_s w_{is}} \frac{w_{js}}{\sum_k w_{ks}}.\label{eq:lipschitzLinearFBound}
\end{align}
We conclude that $g_i(e)=L\cdot e~\forall i$ achieves the maximum over $L$-Lipschitz functions $g$ in Equation \eqref{eqn:lipschitzMinimaxStatement}. Returning to that statement, we are now able to bound the bias using Equation \eqref{eq:lipschitzLinearFBound} as

\begin{align*}
    \arg\min_\mathcal{C} \max_{\{g_i\in\text{Lip}_L(e)\}} \left\vert\E[\hat\tau] - \tau^* \right\vert
    &= \arg\min_\mathcal{C}
    \frac{2}{N} \frac{K}{K-1} L \sum_{i\in[N]} \sum_{j\not\in\mathcal{C}(i)} \sum_{s} \frac{w_{is}}{\sum_s w_{is}} \frac{w_{js}}{\sum_k w_{ks}}.
\end{align*}
Since $K$, $N$ and $L$ are constants with respect to the clustering $\mathcal{C}$, we recover the minimax optimality of the folded graph clustering:
\begin{align*}
    \arg\min_\mathcal{C} \max_{\{g_i\in\text{Lip}_L(e)\}} \left\vert\E[\hat\tau] - \tau^* \right\vert
    &= \arg\min_\mathcal{C} H(\mathcal{C})
\end{align*}

\subsection{Proof of Lemma \ref{lem:exposureMapping} (bounding the bias under the $\Delta-$neighborhood potential outcomes model)}

We begin by applying Lemma \ref{lem:unitLevelBiasDecomp} to decompose the bias in terms of its unit-level contributions:
{\small
\begin{align*}
    \E[\hat\tau] - \tau^* &= \frac{1}{N} \sum_{i\in [N]} \E\left[g_i(Z_i, e_i) - g(Z_i,Z_i)\vert Z_i = 1\right] - \frac{1}{N} \sum_{i\in[N]} \E\left[g_i(Z_i, e_i) - g(Z_i, Z_i) \vert Z_i = -1\right].
\end{align*}
}
Next, we apply the property of the potential outcomes given in \eqref{eqn:exposureMappingInterference}. We know from the assumption that whenever $\vert Z - e\vert < \Delta$, we have $\vert g_i(Z,e) - g_i(Z,Z)\vert =0$, and otherwise we have $\vert g_i(Z,e) - g_i(Z,Z)\vert \leq B$. We can therefore bound the absolute value of the bias by $m$ times the probability that each unit's exposure $e_i$ deviates by more than $\Delta$ from its assignment $Z_i$.
\begin{align*}
    \vert \E[\hat\tau] - \tau^* \vert 
    &\leq \frac{1}{N} \sum_{i\in [N]} \E\left[B \mathbf{1}\{ e_i < 1 - \Delta \}\vert Z_i = 1\right] - \frac{1}{N} \sum_{i\in[N]} \E\left[B \mathbf{1}\{ e_i > -1 + \Delta\} \vert Z_i = -1\right]\\
    &= \frac{1}{N} \sum_{i\in [N]} B\P\left(e_i < 1 - \Delta \vert Z_i = 1\right) - \frac{1}{N} \sum_{i\in[N]} B\P\left( \mathbf{1}\{ e_i > -1 + \Delta\} \vert Z_i = -1\right)\\
    &= \frac{1}{N} \sum_{i\in [N]} B\P\left(1 - e_i > \Delta \vert Z_i = 1\right) - \frac{1}{N} \sum_{i\in[N]} B\P\left( 1 + e_i > \Delta \vert Z_i = -1\right)
\end{align*}
We apply Markov's inequality to both terms:
\begin{align*}
    \vert \E[\hat\tau] - \tau^* \vert 
    &\leq \sum_{i\in[N]} \frac{B}{\Delta} \left( \E[1 - e_i \vert Z_i = 1] + \E[1 + e_i  \vert Z_i = -1]\right)
\end{align*}
Next, we apply Lemma \ref{lem:conditionalSumToGraphObjective} with $\gamma_i=1$ to write this expression in terms of our folded graph clustering objective, which completes the proof.
\begin{align*}
    \vert \E[\hat\tau] - \tau^* \vert 
    &\leq \frac{2B}{N\Delta} \frac{K}{K-1} \sum_{i\in[N]} \sum_{j\not\in\mathcal{C}(i)} \sum_{s} \frac{w_{is}}{\sum_s w_{is}} \frac{w_{js}}{\sum_k w_{ks}}.
\end{align*}
\qed

\subsection{Proof of Lemma \ref{lem:covarianceObjective} (The objective $\mathcal{H}(\mathcal{C})$ maximizes the covariance between exposure and treatment assignment)}

In this proof, we will use the linearity of $\mathbf{e}$ in $\mathbf{Z}$ to write the covariance objective entirely in terms of linear combinations of $Z_i$. We will then use the fact that $\mathcal{D}(\mathcal{C})$ is a cluster-randomized design to compute the covariance exactly, and show that minimizing the covariance objective is identical to minimizing the folded graph clustering objective.

We begin by rewriting our optimization objective in terms of only the treatment assignments $Z$. Recall that, under the linear dose and exposure mappings, Lemma \ref{lem:e=CZ}, $\mathbf{e}=C\mathbf{Z}$ for a known matrix $C$ that depends only on the interference graph.
\begin{align*}
    \arg\max_{\mathcal{C}} ~Tr\left( Cov_{Z\sim\mathcal{D}(\mathcal{C})}(\mathbf{Z}, \mathbf{e}) \right)
    &= \arg\max_{\mathcal{C}} ~Tr\left(\E_{\mathbf{Z}\sim \mathcal{D}(\mathcal{C})}\left[ (\mathbf{Z} - \E[\mathbf{Z}]) (\mathbf{e} - \E[\mathbf{e}])^T\right] \right)\\
    &= \arg\max_{\mathcal{C}} ~Tr\left(\E_{\mathbf{Z}\sim \mathcal{D}(\mathcal{C})}\left[ (\mathbf{Z} - \E[\mathbf{Z}]) (C\mathbf{Z} - \E[C\mathbf{Z}])^T\right] \right)\\
    &= \arg\max_{\mathcal{C}} ~\E_{\mathbf{Z}\sim \mathcal{D}(\mathcal{C})}\left[ Tr\left(C(\mathbf{Z} - \E[\mathbf{Z}]) (\mathbf{Z} - \E[\mathbf{Z}])^T \right) \right]
\end{align*}
Observe that we are taking the trace of a product of two $N\times N$ matrices, $C$ and $(\mathbf{Z} - \E[\mathbf{Z}]) (\mathbf{Z} - \E[\mathbf{Z}])^T$. The trace of a product of square matrices is the sum of the entries in their elementwise (Hadamard) product, which lets us write the trace as a sum and apply linearity of expectation:
\begin{align}
    \arg\max_{\mathcal{C}} ~Tr\left( Cov_{\mathbf{Z}\sim\mathcal{D}(\mathcal{C})}(\mathbf{e}, \mathbf{Z}) \right)
    &= \arg\max_{\mathcal{C}} ~\E_{\mathbf{Z}\sim \mathcal{D}(\mathcal{C})}\left[ \sum_{i,j\in [N]} (Z_i - \E[Z_i])(Z_j - \E[Z_j]) C_{ij}\right]\notag\\
    &= \arg\max_{\mathcal{C}} ~\sum_{i,j\in [N]} C_{ij} \E_{\mathbf{Z}\sim \mathcal{D}(\mathcal{C})}\left[ (Z_i - \E[Z_i])(Z_j - \E[Z_j]) \right] \label{eqn:covarianceBeforeClustering}
\end{align}
We have by Definition \ref{def:balancedDesign} that $\mathcal{D}(\mathcal{C})$ is a uniform assignment of $K$ balanced clusters into $K_T$ treated units and $K_C=K-K_T$ control units. We see that the value of the expectation $\E_{\mathbf{Z}\sim \mathcal{D}(\mathcal{C})}\left[ (Z_i - \E[Z_i])(Z_j - \E[Z_j]) \right] $ depends on whether units $i$ and $j$ belong to the same or different clusters. We have
\begin{align*}
    \arg\max_{\mathcal{C}} ~Tr&\left( Cov_{Z\sim\mathcal{D}(\mathcal{C})}(\mathbf{e}, \mathbf{Z}) \right)=\\
    & \arg\max_{\mathcal{C}} ~\sum_{i} \Big( \sum_{j\in \mathcal{C}(i)} C_{ij} \E_{\mathbf{Z}\sim \mathcal{D}(\mathcal{C})}\left[ (Z_i - \E[Z_i])(Z_j - \E[Z_j]) \vert j\in \mathcal{C}(i)\right] \\
    &\qquad\qquad\qquad\qquad + \sum_{j\not\in\mathcal{C}(i)} C_{ij} \E_{Z\sim \mathcal{D}}\left[ (Z_i - \E[Z_i])(Z_j - \E[Z_j]) \vert j\not\in \mathcal{C}(i) \right] \Big)
\end{align*}
Our next step will be to compute both conditional expectations. We will see that the conditional expectations are independent of the indices $i$ and $j$ (since the argument of the expectation depends only on whether $i$ and $j$ belong to the same cluster), and that the conditional expectation is greater when $i$ and $j$ belong to different clusters. This will let us draw an equivalence between our covariance objective and the objective of minimizing the folded graph cut, which will turn out to be exactly the objective of minimizing cuts in $C_{ij}$. We expand the conditional covariances and use linearity of expectation, along with the fact that $\E[Z_j]=(K_T-K_C)/K$ to write
\begin{align*}
    \E_{\mathbf{Z}\sim \mathcal{D}(\mathcal{C})}&\left[ (Z_i - \E[Z_i])(Z_j - \E[Z_j]) \vert j\in \mathcal{C}(i)\right] \\
    &= \E_{\mathbf{Z}\sim \mathcal{D}(\mathcal{C})}\left[ Z_i Z_j \vert j\in \mathcal{C}(i) \right] -
    \E_{\mathbf{Z}\sim \mathcal{D}(\mathcal{C})}\left[ Z_i \vert j\in \mathcal{C}(i) \right] \E[Z_j] \\&\qquad-  \E_{Z\sim \mathcal{D}(\mathcal{C})}\left[ Z_j \vert j\in \mathcal{C}(i) \right] \E[Z_i] + 
   \E[Z_j] \E[Z_i] \\
    &= \E_{\mathbf{Z}\sim \mathcal{D}(\mathcal{C})}\left[ Z_i Z_j \vert j\in \mathcal{C}(i) \right] -
    \E\left[ Z_i \right] \E[Z_j] - 
    \E\left[ Z_j \right] \E[Z_i] + 
    \E[Z_j] \E[Z_i] \\
    &= \E_{\mathbf{Z}\sim \mathcal{D}(\mathcal{C})}\left[ Z_i Z_j \vert j\in \mathcal{C}(i) \right] -
    \E[Z_j] \E[Z_i] \\
    &= \E_{\mathbf{Z}\sim \mathcal{D}(\mathcal{C})}\left[ Z_i Z_j \vert j\in \mathcal{C}(i) \right] -
    \left( \frac{K_T - K_C}{K} \right)^2\\
    &= 1 -
    \left( \frac{K_T - K_C}{K} \right)^2
\end{align*}
and similarly
\begin{align*}
    \E_{\mathbf{Z}\sim \mathcal{D}(\mathcal{C})}&\left[ (Z_i - \E[Z_i])(Z_j - \E[Z_j]) \vert j\not\in \mathcal{C}(i)\right]\\
    &= \E_{\mathbf{Z}\sim \mathcal{D}(\mathcal{C})}\left[ Z_i Z_j \vert j\not\in \mathcal{C}(i) \right] - \E[Z_j] \E[Z_i] \\
    &= 1\cdot\frac{K_T}{K}\cdot\frac{K_T-1}{K-1} + (-1)\cdot\frac{K_T}{K}\frac{K_C-1}{K-1} + (-1)\cdot\frac{K_C}{K}\frac{K_T-1}{K-1} \\&\qquad+ 1\cdot\frac{K_C}{K}\frac{K_C-1}{K-1} - \E[Z_j] \E[Z_i]\\
    &= \frac{K_T(K_T - 1) - K_T (K_C - 1) - K_C (K_T - 1) + K_C (K_C - 1)}{K(K-1)} - \E[Z_j] \E[Z_i]\\
    &= \frac{K_T(K_T - 1) - K_T (K_C - 1) - K_C (K_T - 1) + K_C (K_C - 1)}{K(K-1)} - \left(\frac{K_T - K_C}{K} \right)^2\\
    &= \frac{(K_T - K_C)^2 }{K(K-1)} - \left(\frac{K_T - K_C}{K} \right)^2
\end{align*}
Substituting these conditional expectations, we have
\begin{align*}
    \arg\max_{\mathcal{C}} ~Tr&\left( Cov_{\mathbf{Z}\sim\mathcal{D}(\mathcal{C})}(\mathbf{e}, \mathbf{Z}) \right)
    = \\&\arg\max_{\mathcal{C}} ~\sum_{i} \left( \sum_{j\in [N]} C_{ij} \left(1 - \left(\frac{K_T - K_C}{K}\right)^2 \right)
    + \sum_{j\not\in \mathcal{C}(i)} C_{ij} \left(\frac{(K_T - K_C)^2 }{K(K-1)} - 1\right)
    \right)
\end{align*}
Next, we recognize that the sum over all $i$ and $j$ is constant with respect to the cluster randomized design $\mathcal{C}$ and therefore can be removed from the argmax.
\begin{align*}
    \arg\max_{\mathcal{C}} ~Tr\left( Cov_{\mathbf{Z}\sim\mathcal{D}(\mathcal{C})}(\mathbf{e}, \mathbf{Z}) \right)
    &= \arg\max_{\mathcal{C}} ~\sum_{i} \left( \sum_{j\not\in \mathcal{C}(i)} C_{ij} \left(\frac{(K_T - K_C)^2 }{K(K-1)} - 1\right)
    \right)\\
    &= \arg\min_{\mathcal{C}} ~\sum_{i} \left( \sum_{j\not\in \mathcal{C}(i)} C_{ij} \left(1 - \frac{(K_T - K_C)^2 }{K(K-1)}\right)
    \right)
\end{align*}
Observe that, as long as there is at least one treated and control unit (i.e., $0 < K_T < K$) then the quantity $1 - \frac{(K_T - K_C)^2 }{K(K-1)}$ is positive and can be taken out of the argmax. In this case, we have
\begin{align*}
 \arg\max_{\mathcal{C}} ~Tr\left( Cov_{\mathbf{Z}\sim\mathcal{D}(\mathcal{C})}(\mathbf{e}, \mathbf{Z}) \right)
    &= \arg\min_{\mathcal{C}} ~\sum_{i} \left( \sum_{j\not\in \mathcal{C}(i)} C_{ij}
    \right)
\end{align*}
Finally, we substitute the value of $C_{ij}$ from Lemma \ref{lem:e=CZ}:
\begin{align*}
     \arg\max_{\mathcal{C}} ~Tr\left( Cov_{\mathbf{Z}\sim\mathcal{D}(\mathcal{C})}(\mathbf{e}, \mathbf{Z}) \right)
    &= \arg\min_{\mathcal{C}} ~\sum_{i} \left( \sum_{j\not\in \mathcal{C}(i)} 
    \frac{\sum_s w_{is} \frac{w_{js}}{\sum_k w_{ks}}}{\sum_s w_{is}}
    \right)
\end{align*}
which we recognize as precisely the minimum-cut objective on the folded graph.
\qed
\section{Standard deviation and RMSE for experiments}

We set up the parameters of our first two experiments so that the error of $\widehat\tau_{DIM}$ was almost entirely due to bias, instead of variance. Here we provide the normalized standard deviation and RMSE for the experiments in Sections \ref{sec:experiments:robustness-to-graph} and \ref{sec:experiments:nonlinearity}, for completeness.

\subsection{Robustness to different graph structures}

See Tables \ref{tab:robustness-to-graph-rmse} and \ref{tab:robustness-to-graph-std}.

\begin{table}[h]
    \centering
    \caption{Relative RMSE of $\widehat\tau_{DIM}$ as the bipartite stochastic block model changes (see \ref{sec:experiments:robustness-to-graph})\newline}
    \label{tab:robustness-to-graph-rmse}
    \begin{tabular}{rrrrr} 
    \toprule
    & $p=0.0$ & $p=0.005$ & $p=0.05$ & $p=0.5$\\
    \hline 
    $\mathcal{H}(\mathcal{C})$&0.23($\pm$0.03)&3.84($\pm$0.05)&11.54($\pm$0.06)&12.99($\pm$0.09)\\
    $\text{Tr}(\text{Var}(\mathbf{d}))$&0.28($\pm$0.03)&3.86($\pm$0.05)&11.49($\pm$0.06)&12.92($\pm$0.07)\\
    Direct clustering&0.26($\pm$0.03)&9.26($\pm$0.14)&12.69($\pm$0.07)&12.95($\pm$0.07)\\
    EXPOSURE-DESIGN&0.53($\pm$0.05)&4.06($\pm$0.07)&11.9($\pm$0.07)&13.0($\pm$0.08)\\
    Unit-level randomization&12.43($\pm$0.08)&12.55($\pm$0.09)&12.77($\pm$0.07)&12.96($\pm$0.08)\\
    \midrule
    True clusters&0.31($\pm$0.04)&3.86($\pm$0.04)&11.58($\pm$0.06)&12.96($\pm$0.06)\\
    \bottomrule
    \end{tabular}
\end{table}

\begin{table}[h]
    \centering
    \caption{Relative standard deviation of $\widehat\tau_{DIM}$ as the bipartite stochastic block model changes (see \ref{sec:experiments:robustness-to-graph})\newline}
    \label{tab:robustness-to-graph-std}
    \begin{tabular}{rrrrr} 
    \toprule
    & $p=0.0$ & $p=0.005$ & $p=0.05$ & $p=0.5$\\ 
    \hline 
    $\mathcal{H}(\mathcal{C})$&0.23($\pm$0.03)&0.26($\pm$0.04)&0.28($\pm$0.04)&0.42($\pm$0.05)\\
    $\text{Tr}(\text{Var}(\mathbf{d}))$ &0.28($\pm$0.04)&0.28($\pm$0.03)&0.31($\pm$0.05)&0.37($\pm$0.05)\\
    Direct clustering&0.26($\pm$0.03)&0.69($\pm$0.09)&0.37($\pm$0.05)&0.35($\pm$0.05)\\
    EXPOSURE-DESIGN&0.37($\pm$0.04)&0.33($\pm$0.04)&0.33($\pm$0.04)&0.41($\pm$0.05)\\
    Unit-level Randomization&0.39($\pm$0.05)&0.45($\pm$0.07)&0.38($\pm$0.05)&0.41($\pm$0.06)\\
    \midrule
    True Clusters&0.31($\pm$0.03)&0.23($\pm$0.03)&0.28($\pm$0.03)&0.29($\pm$0.04)\\
    \bottomrule
    \end{tabular}
\end{table}

\subsection{Robustness to nonlinearity}

See Tables \ref{tab:robustness-to-linearity-rmse} and \ref{tab:robustness-to-linearity-std}.

\begin{table}[h]
    \caption{Relative RMSE of $\widehat\tau_{DIM}$ as the neighborhood of pure exposure, $\Delta$, widens (see \ref{sec:experiments:nonlinearity})\newline} 
    \label{tab:robustness-to-linearity-rmse}
    \centering
    \begin{tabular}{rccc} 
    \toprule
    & $\Delta= 0.1$ & $\Delta=0.3$ & $\Delta=0.5$\\
    \midrule
     $\mathcal{H}(\mathcal{C})$&1.0($\pm$0.004)&0.458($\pm$0.005)&0.001($\pm$0.0)\\
     $\text{Tr}(\text{Var}(\mathbf{d}))$&1.002($\pm$0.004)&0.461($\pm$0.004)&0.001($\pm$0.0)\\
     Direct clustering&0.997($\pm$0.005)&0.95($\pm$0.008)&0.608($\pm$0.02)\\
     EXPOSURE-DESIGN&1.001($\pm$0.004)&0.509($\pm$0.005)&0.011($\pm$0.002)\\
     Unit-level randomization&0.998($\pm$0.004)&1.0($\pm$0.004)&0.998($\pm$0.003)\\
     \midrule
     True clusters&1.001($\pm$0.004)&0.459($\pm$0.004)&0.001($\pm$0.0)\\
    \bottomrule
    \end{tabular}
\end{table}

\begin{table}[h]
    \caption{Relative standard deviation of $\widehat\tau_{DIM}$ as the neighborhood of pure exposure, $\Delta$, widens (see \ref{sec:experiments:nonlinearity})\newline} 
    \label{tab:robustness-to-linearity-std}
    \centering
    \begin{tabular}{rccc} 
    \toprule
    & $\Delta= 0.1$ & $\Delta=0.3$ & $\Delta=0.5$\\
    \midrule
    $\mathcal{H}(\mathcal{C})$&0.02($\pm$0.003)&0.024($\pm$0.003)&0.001($\pm$0.0)\\
    $\text{Tr}(\text{Var}(\mathbf{d}))$&0.02($\pm$0.002)&0.021($\pm$0.003)&0.001($\pm$0.0)\\
    Direct clustering&0.026($\pm$0.003)&0.038($\pm$0.007)&0.097($\pm$0.01)\\
    EXPOSURE-DESIGN&0.02($\pm$0.003)&0.025($\pm$0.003)&0.005($\pm$0.001)\\
    Unit-level randomization&0.019($\pm$0.002)&0.019($\pm$0.002)&0.017($\pm$0.002)\\
    \midrule
    True clusters&0.021($\pm$0.003)&0.024($\pm$0.003)&0.001($\pm$0.0)\\
    \bottomrule
    \end{tabular}
\end{table}
\section{Description of the Balanced Partitioning Algorithm}\label{sec:algorithm}

We provide here a brief overview of the clustering algorithms used in our paper. For each of the $\mathcal{H}(\mathcal{C})$, direct clustering, and $\text{Tr}(\text{Var}(\mathbf{d}))$ objectives, we used an implementation of a balanced partitioning algorithm, kindly provided by the authors of~\citet{aydin2019distributed}, on the appropriately constructed graph.
\begin{itemize}
    \item For the $\mathcal{H}(\mathcal{C})$ objective, the graph includes only the experimental units as nodes. Each node weight is set to 1, and the edge weight between a pair of nodes is given by Equation~\eqref{eq:graphicalInterpretation}.
    \item For the direct clustering, the graph includes both experimental units and interference units as nodes. All experimental units have node weights set to 1, and all interference units have node weights set to 0. The edge weight between them is kept as is. The interference units are removed from the clusters once these are computed in order to produce experiment-unit-only clusters.
    \item For the $\text{Tr}(\text{Var}(\mathbf{d}))$, the graph includes only the experimental units as node. Each node weight is set to 1, and the edge weight between a pair of nodes is given by Equation~\eqref{eqn:TrVarD}.
\end{itemize}

In all balanced partitioning runs, the number of clusters is given as fixed, equal to $10$ unless specified otherwise. The maximum allowed imbalanced (ratio between the largest and the smallest cluster by sum of node weights) is 10\%. The algorithm runs in two steps:
\begin{enumerate}
    \item An initial embedding of nodes is given onto a line with affinity clustering. This ordering is then broken into clusters by taking contiguous equally-weighted segments of the line.
    \item Node swaps are then evaluated to improve the cut size in a post-processing procedure. At most 2 post-processing passes are done before outputting the final clusters.
\end{enumerate}

Please see \citet{aydin2019distributed} for more information. For the EXPOSURE-DESIGN objective, we use the implementation kindly provided by the authors of~\citet{harshaw2022design}. As described in Section 7.3 of that work, the algorithm uses a greedy local search, starting from singleton clusters, to assign units to clusters. The role of ``diversion units'' in their paper plays the role of the experimental units in ours, while the ``outcome units'' in their paper plays the role of the interference units in ours. Their clustering objective in Proposition 7.1 allows for the tuning of a hyper-parameter $\lambda$. We experimented with values $\lambda \in \{0, 0.001, 0.01, 0.1, 1\}$ and reported the best outcome in each case, best defined by minimum mean-squared error unless specified otherwise. To further regularize the output, their implementation adds a regularization term to the objective in the form of $0.95^k \times \sum_{C} W_C^2$, where $W_C^2$ is the current weight of cluster $C$, and $k$ is the number of iterations completed.

\end{document}